\documentclass[fleqn,10pt]{wlscirep}
\usepackage[utf8]{inputenc}
\usepackage[T1]{fontenc}
\usepackage{caption}
\usepackage{float} 
\usepackage{hyperref}

\usepackage{amsmath} 

\title{Decoding the Hidden Dynamics of Super-Arrhenius Hydrogen Diffusion in Multi-Principal Element Alloys via Machine Learning}

\author[1,*]{Fei Shuang}
\author[1,2]{Yucheng Ji}
\author[1]{Zixiong Wei}
\author[2]{Chaofang Dong}
\author[3,4]{Wei Gao}
\author[5]{Luca Laurenti}
\author[1,*]{Poulumi Dey}

\affil[1]{Department of Materials Science and Engineering, Faculty of Mechanical Engineering, Delft University of Technology, Mekelweg 2, Delft, 2628 CD, The Netherlands}
\affil[2]{Beijing Advanced Innovation Center for Materials Genome Engineering, National Materials Corrosion and Protection Data Center, Institute for Advanced Materials and Technology, University of Science and Technology Beijing, Beijing 100083, China}
\affil[3]{J. Mike Walker’66 Department of Mechanical Engineering, Texas A\&M University, College Station, TX 77843, United States}
\affil[4]{Department of Materials Science \& Engineering, Texas A\&M University, College Station, TX 77843, United States}
\affil[5]{Delft Centre of System and Control (DCSC), Faculty of Mechanical Engineering, Delft University of Technology, Mekelweg 2, Delft, 2628 CD, The Netherlands}
\affil[*]{P.dey@tudelft.nl, F.Shuang@tudelft.nl}

\begin{abstract}

Understanding atomic hydrogen (H) diffusion in multi-principal element alloys (MPEAs) is essential for advancing clean energy technologies such as H transport, storage, and nuclear fusion applications. However, the vast compositional space and the intricate chemical environments inherent in MPEAs pose significant obstacles. In this work, we address this challenge by developing a multifaceted machine learning framework that integrates machine-learning force field, neural network-driven kinetic Monte Carlo, and machine-learning symbolic regression. This framework allows for accurate investigation of H diffusion across the entire compositional space of body-centered cubic (BCC) refractory MoNbTaW alloys, achieving density functional theory accuracy. For the first time, we discover that H diffusion in MPEAs exhibits super-Arrhenius behavior, described by the Vogel-Fulcher-Tammann model, where the Vogel temperature correlates with the 5th percentile of H solution energy spectrum. We also derive robust analytical expressions that can be used to predict H diffusivity in general BCC MPEAs. Our findings further elucidate that chemical short-range order (SRO) generally does not impact H diffusion, except it enhances diffusion when “H-favoring” elements (notably Nb and Ta) are present in low concentrations. These findings not only enhance our understanding of H diffusion dynamics in general MPEAs but also guide the development of advanced MPEAs in H-related applications by manipulating element type, composition and SRO.
\end{abstract}
\begin{document}
\flushbottom
\maketitle
\thispagestyle{empty}
\section*{Introduction}
Green energy is the cornerstone of a seismic shift toward sustainable future, dismantling fossil fuel dependencies and spearheading the battle against climate change while revolutionizing eco-friendly energy systems\cite{Cheng2023,Odenweller2022}. Within this paradigm, metals and alloys stand as pivotal catalysts for hydrogen (H) production and guardians of storage system integrity\cite{Raabe2019}. The burgeoning frontier of multi-principal element alloys (MPEA), including high entropy alloys (HEAs) and medium-entropy alloys (MEAs), marks an unparalleled leap in materials science\cite{Ye2016,George2019,Marques2021-1}. Offering an expansive compositional realm, MPEAs harbor unprecedented potential for innovation in H-related applications\cite{Marques2021-1}. For instance, recent breakthroughs have illuminated the exceptional resistance of specific MPEAs, such as face-centered cubic (FCC) equimolar CoCrFeMnNi\cite{Luo2017,Luo2018-1,Luo2018-2}, and CoNiV\cite{Luo2020}, against H embrittlement. A crucial factor contributing to this exceptional resistance is the intricate chemical environments at the atomic scale within these alloys, serving as highly effective sites for H trapping thereby resulting in low H diffusivity. On the other hand, nuclear fusion power systems stand out as another promising solution for sustainable clean energy\cite{Knaster2016}. The exceptionally harsh operational conditions of fusion reactors, including exposure to deuterium-tritium plasma and enduring severe fluxes of particles and heat, coupled with the radiation damage induced by high-energy neutrons, however, pose a substantial challenge for the use of the existing structural materials\cite{Zinkle2009}. Fortunately, recent experiments have unveiled that body-centered cubic (BCC) W-based refractory MPEAs such as WTaCrV\cite{El-Atwani-2019} and WTaCrVHf\cite{El-Atwani-2023} demonstrate remarkable radiation resistance and thermal stability, offering a promising solution to address these challenges. The critical but unclear issue for MPEAs in nuclear fusion applications pertains to the retention and recycling of H isotopes, where the solubility and diffusivity of H isotopes play a crucial role\cite{Causey2002}. Therefore, unraveling the mechanisms governing H diffusion dynamics within MPEAs' diverse compositional spectrum is paramount, heralding a new era in material engineering for sustainable energy solutions. 

In the quest for accurate H diffusion examination, challenges persist due to lattice defects such as grain boundaries, vacancies, impurities, and intricate surface complexities\cite{Kwon2023,Nagumo2023}. Consequently, experimental assessments often inadequately capture the correct diffusion coefficients within the lattice. Notably, the experimentally reported diffusion coefficient variations spanning several orders of magnitude for BCC Fe and W at room temperature highlight the considerable hurdles in accurately characterizing this fundamental property\cite{Frauenfelder1969,Kiuchi1983,Otsuka2009}. Nevertheless, intensive experimental endeavors have aimed to directly gauge or estimate H diffusivity through permeation tests conducted in MPEAs\cite{Belo2020,Zhu2023,Suzuki2015,Zhu2019}. However, exploring diffusion mechanisms at the atomic scale and quantitatively analyzing the influence of chemical environments on H diffusion remain challenging due to the limitations in experimental resolution. Another extensively utilized approach involves integrating density functional theory (DFT) calculations with kinetic Monte Carlo (KMC) simulations. This method is commonly employed to explore H diffusion processes within lattices and defects in conventional metals and alloys\cite{Boda2020,Samin2019,Zhang2023,Zhou2012}. However, the comprehensive calculation of every diffusion barrier within the myriad chemical environments of MPEAs across all compositions remains unattainable.

The advent of machine learning (ML) and machine learning force field (MLFF) have empowered insightful studies of H diffusion in pure metals, alloys, and MPEAs\cite{Liu2023,Tang2024}. For example, several ML models have been trained to predict H solution energies at critical sites within FCC CoCrFeMnNi and its subsystems, while KMC simulations were utilized to compute H diffusion coefficients\cite{Zhou2022}. Yet, these ML models solely provide estimates of critical energies with substantial errors, falling short of an accurate force field necessary for reliable investigation of H diffusion. Another interesting study utilized reinforcement learning to guide long-timescale simulations of H diffusion in a medium-entropy CrCoNi alloy \cite{Tang2024}, employing a universal neural network-based interatomic potential \cite{Takamoto2022} to accurately calculate the energy landscape. The use of this universal potential raises concerns about the accuracy of the energy calculations, as its broad applicability may not capture the specific interactions unique to this alloy system. Recently, quantum-accurate MLFFs have emerged for investigating H diffusion in BCC Fe, Nb, and W\cite{Meng2021,Kwon2023}. However, developing a reliable MLFF for H diffusion in MPEAs faces substantial challenges when applying a similar approach. Examining a simple configuration with four alloying elements in a 5 \text{\AA} range (comprising about 30 neighboring atoms) reveals approximately $4^{30}$ potential H-metal environments. Furthermore, the small size of H atom enables its occupation across numerous sites within the metallic matrix. Moreover, atomic size mismatch and thermal fluctuations induce substantial lattice distortions among metallic atoms, further complicating the H-metal environments. While advanced active learning strategies have facilitated the development of various MLFFs\cite{Jinnouchi2020,Novikov2021}, performing long-timescale simulations (lasting hundreds of nanoseconds) in ab initio molecular dynamics (MD) still remains challenging. The integration of MLFF-based MD with DFT calculations, essential for on-the-fly active learning, also encounters significant obstacles. These challenges are even more pronounced in diffusion studies at low temperatures, where the limitations of MD simulations in covering extended timescales are most apparent\cite{Qi2024}.

In this study, we present an advanced machine learning computational framework to investigate the complex dynamics of H diffusion in MoNbTaW MPEAs. While the mechanical properties of these MPEAs have been extensively studied \cite{Zou2014,Zou2015,Yin2021,Pedro2023}, their H diffusivity and solubility remain unexplored. To address this gap, we have developed a highly accurate MLFF for MoNbTaW-H systems. This MLFF is constructed using a D-optimality-based pre-selection (DPS) approach to identify representative configurations before performing DFT calculations, offering a significant departure from traditional active learning methods. Furthermore, we introduce a neural network (NN) model that predicts diffusion barriers using lattice distortion-corrected atomic descriptors. This model forms the foundation of our scalable neural network-driven Kinetic Monte Carlo (NN-KMC) framework, which facilitates an in-depth exploration of H diffusion dynamics across the entire compositional range of MoNbTaW. Another key innovation in our work is the use of a constrained KMC method, allowing us to isolate the effects of the complex energy landscape from those of temperature on H diffusion, thus providing a detailed analysis of the super-Arrhenius behavior observed in MPEAs. We also employ machine-learning symbolic regression (MLSR) to develop interpretable models that provide physical insights into H diffusion dynamics. Additionally, we utilize the NN-KMC to examine the impact of chemical short-range order (SRO) on H diffusivity, revealing the specific mechanisms that govern H diffusion.

\section*{Results}

\subsection*{Machine learning-driven H diffusion simulations at DFT-Level Accuracy}
\begin{figure}[!ht]
    \centering
    \includegraphics[width=1\linewidth]{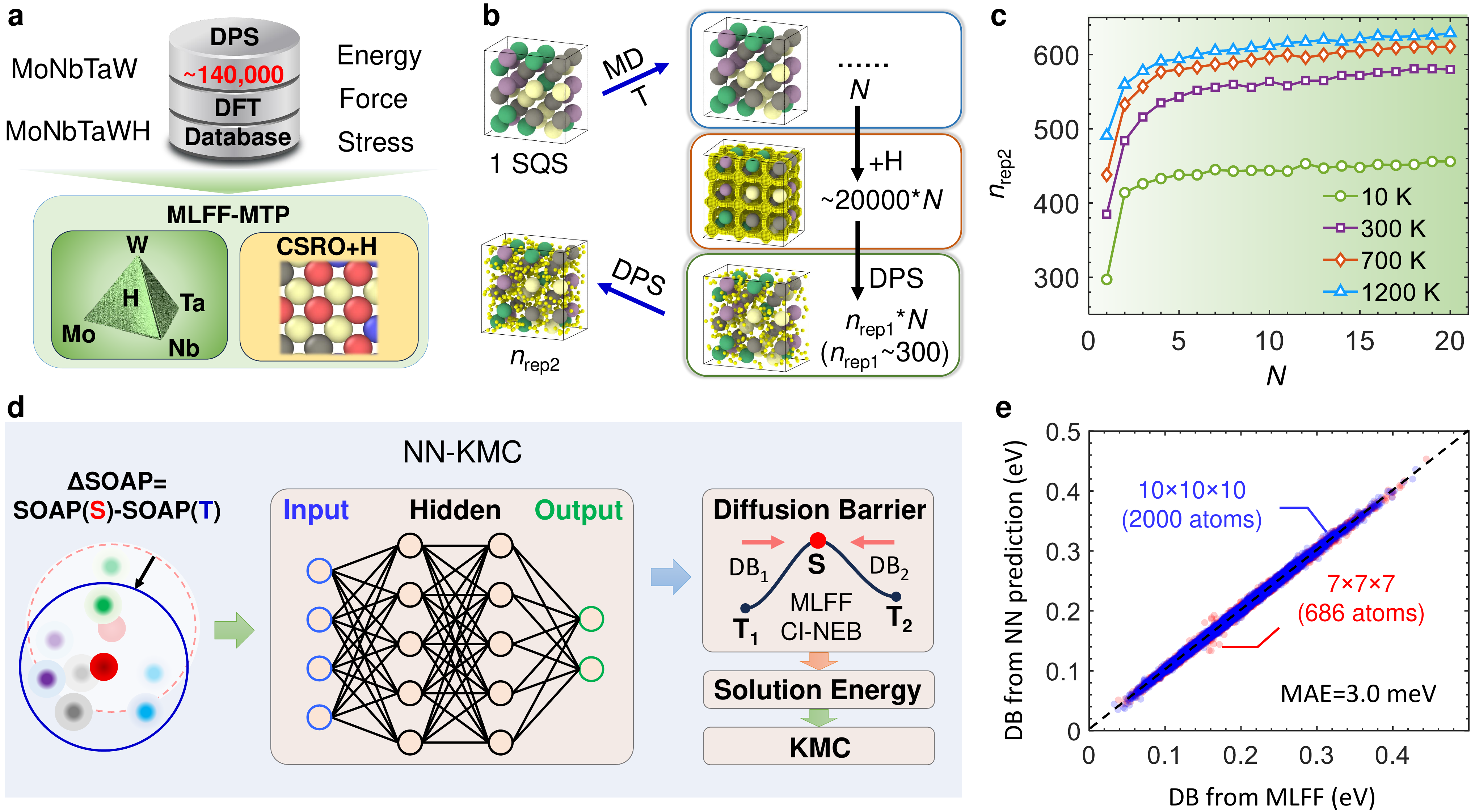}
    \caption{\textbf{Machine learning-driven exploration of H diffusion at DFT-Level Accuracy.} \textbf{a} Flowchart of MLFF development. \textbf{b} Schematic of DPS at various temperatures. $n_\text{rep1}$ and $n_\text{rep2}$ indicate the chosen configurations following the initial and subsequent DPS operations, while $N$ signifies the number of distorted MPEA configurations at a temperature $T$. Large spheres with different colors represent metallic atoms, and small yellow spheres represents H atoms. \textbf{c} The progression of representative configurations ($n_\text{rep2}$) following a two-level DPS process with the count of distorted MPEA configurations ($N$) at varying temperatures. \textbf{d} Flowchart of neural network-driven kinetic Monte Carlo (NN-KMC). \textbf{e} Validation of neural network model scalability for barrier predictions in $7 \times 7 \times 7$ and $10 \times 10 \times 10$ supercells.}
    \label{fig1}
\end{figure}

We have developed a comprehensive machine learning computational framework to explore H diffusion dynamics in MPEAs with an accuracy comparable to DFT. This framework consists of two pivotal components, each detailed in Fig. \ref{fig1}. First, we craft a highly accurate MLFF for MoNbTaW-H system, as shown in Figs \ref{fig1}a-c. Secondly, we establish a scalable NN model capable of predicting H diffusion barriers and solution energies (hereafter abbreviated as DB and SE, respectively), which will be used in KMC simulations for H diffusion (Fig. \ref{fig1}d).  

In developing the MLFF for the MoNbTaW-H system, we construct an extensive database encompassing approximately 140,000 configurations with diverse metal-metal and H-metal interactions across the full compositional range of MoNbTaW MPEAs, as illustrated in Fig. \ref{fig1}a. A crucial aspect of our MLFF development is the database construction employing a DPS strategy prior to any DFT calculations (see “Methods” for the technical details of DPS). Fig. \ref{fig1}b illustrates the workflow of the DPS method on a 36-atom $\text{Mo}_{25}\text{Nb}_{25}\text{Ta}_{25}\text{W}_{25}$ special quasi-random structure (SQS). To consider the effects of lattice thermal vibrations of MPEAs during DPS, we sample $N$ distorted configurations at 10 ps intervals from MD simulations conducted at temperature $T$. For each configuration, we identify approximately 20,000 potential H occupation sites using a grid spacing of 0.2 \AA, selecting sites where the H-to-metal distance is greater than 1.75 \AA. The initial DPS operation selects approximately \(n_{\text{rep1}} = 300\) representative configurations from a pool of 20,000. A subsequent DPS operation further reduces the number of required DFT calculations from \(n_{\text{rep1}} \times N\) to \(n_{\text{rep2}}\). Fig. \ref{fig1}c illustrates that \(n_\text{rep2}\) demonstrates an upward trend with both the number of distorted configuration (\(N\)) and temperature (\(T\)). Remarkably, \(n_\text{rep2}\) saturates when \(N\) > 10, indicating a threshold in data requirements. Specifically, DFT calculations for just 629 configurations at a temperature of 1200 K are found to be sufficient to encompass all H-metal interactions for a specific SQS at this temperature. Four different temperatures, 10 K, 300 K, 700 K and 1200 K, are considered in DPS for each SQS. The construction process of the complete database is detailed in Fig. S1 and “Methods”. 

The energies, forces, and stresses of all configurations obtained from DFT calculations are utilized to train a MLFF using the framework of the moment tensor potential (MTP)\cite{Novikov2021}. The training/test mean absolute errors (MAEs) for energy, force, and stress are obtained as 1.21/1.20 meV/atom, 22.78/22.85 eV/\AA, and 47.45/47.53 MPa (Fig. S2a-c), respectively, surpassing the performance of most existing MLFFs\cite{Zuo2020}. The rigorous validation results in Figs S2d-g confirm the broad applicability of our developed MLFF for accurately modeling H-MPEA interactions across the entire compositional space (Fig. S2d) and the SRO effect (Fig. S2e). Furthermore, our MLFF has been demonstrated to accurately calculate DBs in climbing image nudged elastic band (CI-NEB) calculations (Fig. S2f and S2g) and is applicable to H diffusion simulations at the high temperatures of 1500 K and 2000 K (Fig. S3).

Although the MLFF introduced herein offers a significant speed advantage over DFT, it faces substantial challenges in probing H diffusion across extended time scales within the intricate energy landscapes of MPEAs\cite{Tang2024}. To address these challenges, we develop a scalable NN-KMC method that enhances the efficiency and reliability of H diffusion simulations. Fig. \ref{fig1}d delineates the NN-KMC framework, showcasing the input data, NN architecture and outputs. For DB prediction, we utilize our MLFF to conduct high-throughput CI-NEB calculations for all H diffusion paths in MPEA SQSs, covering 500 random compositions of the entire compositional space (969 compositions in total with a concentration interval of 6.25\%). Each SQS has a size of 4 × 4 × 4, totaling 128 metallic atoms and 1536 unique diffusion paths. This yielded a total of 768,000 minimum energy paths. The Smooth Overlap of Atomic Positions (SOAP) descriptor\cite{SOAP-1} is utilized to quantify the metallic environment surrounding one H atom. We represent the diffusion paths using the difference in SOAP vectors between the initial tetrahedral (T) site and saddle (S) point, denoted as $\Delta \text{SOAP} = \text{SOAP}(\text{S}) - \text{SOAP}(\text{T})$. It should be noted that the exact initial T site and saddle point are unknown due to the significant lattice distortion in MPEAs. To tackle this issue, we integrate lattice distortion effects into the SOAP features (see Supplementary Note 2). Figs S4 and S5 compare the performance of our NN model with and without lattice distortion correction. The correction significantly enhances accuracy of the NN model, reducing the MAE from 3.8 meV (without distortion) to 2.1 meV (with distortion).

Through experimentation with different configurations of layers and neuron number (Fig. S4), we ascertain that a NN model with four hidden layers, each containing 32 neurons, is optimal, achieving a MAE of 2.2 meV. To validate the scalability of the developed NN model, we utilize the MLFF and CI-NEB to calculate DB across the full compositions using larger supercells of sizes $7 \times 7 \times 7$ and $10 \times 10 \times 10$, which comprise 686 and 2000 atoms, respectively. Fig. \ref{fig1}e demonstrates that our NN model accurately predicts all forward and backward DBs with a MAE of 3.0 meV, showcasing the high scalability of our NN model for DB predictions. Additionally, our convergence tests confirm that the size $20 \times 20 \times 20$ adequately captures the required complexity for random MPEAs (Fig. S7). Utilizing the NN model, we compute DB spectra for 287 compositions, including all compositions with a concentration interval of 10\% and all equimolar compositions, using large supercells of size \(20 \times 20 \times 20\) (16,000 atoms). This yields a total of 110,208,000 DBs obtained within approximately 1,500 CPU hours. This staggering number of calculations demonstrates the remarkable efficiency of our NN model, significantly accelerating the process compared to traditional CI-NEB, which would require prohibitively long computational time for such a number. Fig. S8a clearly indicates that the rule of mixing fails to accurately predict DB, with deviations increasing in correlation with the standard deviation of DBs. Additionally, we analyze the DB distributions for specific compositions, such as Mo$_{25}$Nb$_{25}$Ta$_{25}$W$_{25}$ and Ta$_{90}$W$_{10}$ (see Fig. S8b). The former exhibits a Gaussian-like distribution, whereas the latter is notably concentrated around lower barriers, reflecting the diverse random energy landscapes inherent to different compositions. This diversity underscores the critical need for developing a highly accurate MLFF to effectively capture these intricate variations. It is important to highlight that our model is also capable of predicting the relative SE for all T sites within the supercell by analyzing the forward and backward DBs. Here, “relative” refers to the comparison against a reference T site with an SE of zero. Specifically, we initialize the SE for the first T site as $\text{SE}_1$ = 0, then the SE for any connected T sites can be determined using the difference in DB, expressed as $\text{SE}_2 = \text{DB}_1 - \text{DB}_2$, where $\text{DB}_1$ and $\text{DB}_2$ are the backward and forward DBs, respectively.

\subsection*{Super-Arrhenius H diffusion dynamics in MPEAs}
We apply the developed NN-KMC to compute H diffusion coefficients across the entire compositional space of MoNbTaW systems. In order to disentangle the temperature effect from chemical heterogeneity in MPEAs on H diffusion, we introduce a constrained kinetic Monte Carlo (cKMC) method, building upon the regular kinetic Monte Carlo (rKMC) framework. At a selected temperature, rKMC is initially executed to record the H trajectory, which is sufficient to calculate a reliable diffusion coefficient. Subsequently, cKMC is employed, forcing H to follow the pre-established trajectory from rKMC, but using varying temperatures to calculate time scale for all jumps. The time histories required for this consistent H trajectory at different temperatures constitute the outcomes of the cKMC method. For pure metals, results from rKMC and cKMC are identical because temperature is the only influencing factor (Fig. S9), which however are not the case in MPEAs due to the intricate chemical environments. H diffusion simulations for all 287 compositions are conducted using both rKMC and cKMC at a wide temperature range from 500 K to 5000 K. Temperatures above 3000 K are utilized to investigate diffusion dynamics under extreme conditions, although these temperatures often exceed the experimental feasibility by potentially surpassing the melting points of MPEAs. We calculate the diffusion coefficients by using:
\begin{equation} \label{eq1}
    D = \frac{\text{MSD}}{6t}
\end{equation}
where MSD is the mean squared displacement for one single H atom, and $t$ is the elapsed time. To ensure consistent accuracy of diffusion coefficient ($D$) calculations across various compositions and temperatures, we require all KMC simulations to continue until MSD reaches $1 \times 10^5 \, \text{\AA}^2$. It should be noted that determining diffusion coefficients in MPEAs presents a challenging task, due to the complex chemical environments that result in non-uniform diffusion. To obtain reliable and consistent diffusion coefficients across various compositions, we propose a distance-based criterion to extract the accurate MSD of a H diffusion trajectory, as discussed in Supplementary Note 3. This approach enhances the accuracy of our diffusion coefficient calculations and provides deeper insight into the H mechanisms within MPEAs.

\begin{figure}[ht]
    \centering
    \includegraphics[width=1\linewidth]{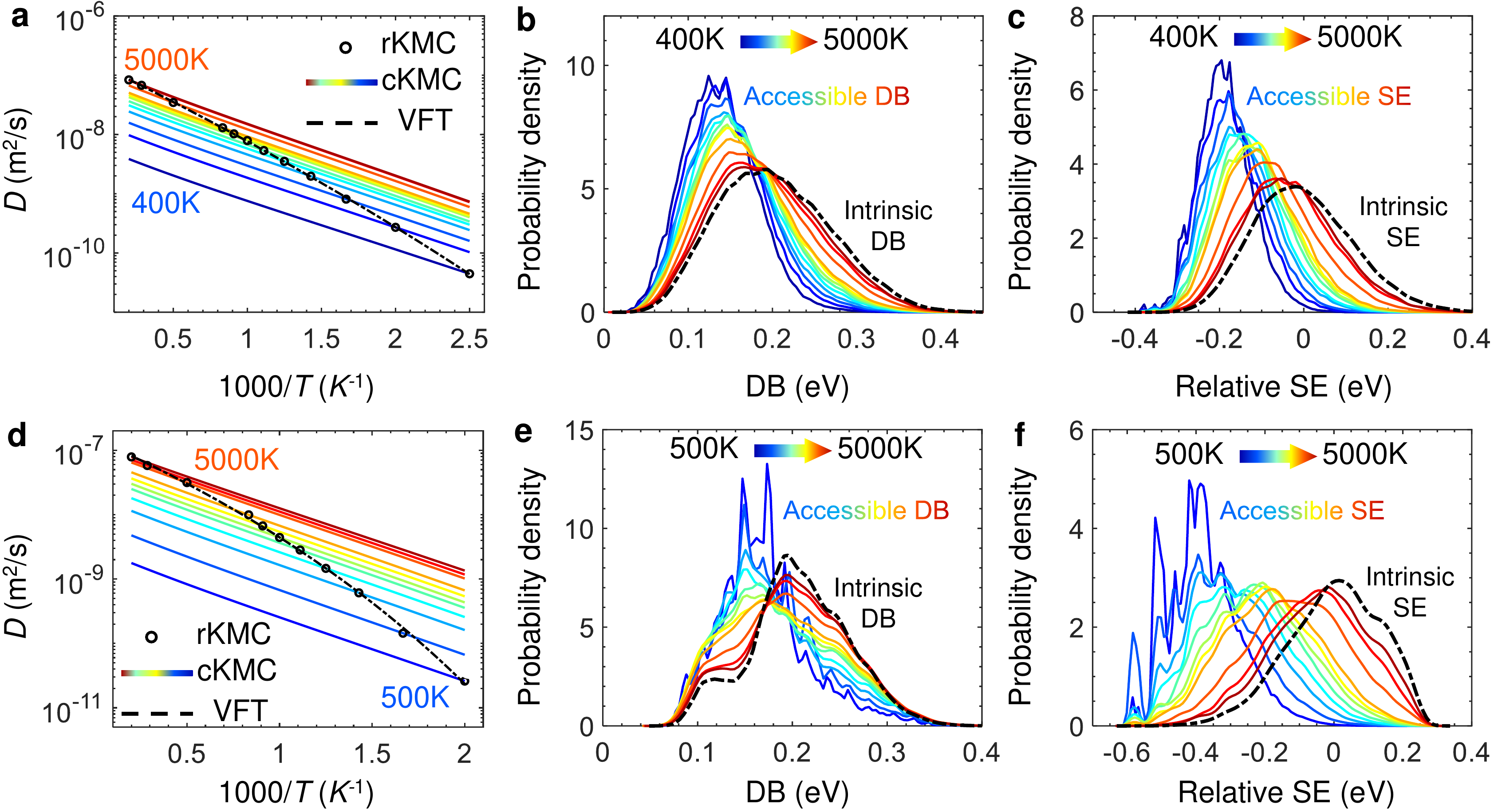}
    \caption{\textbf{Super-Arrhenius H diffusion in $\text{Mo}_{25}\text{Nb}_{25}\text{Ta}_{25}\text{W}_{25}$ and $\text{Ta}_{20}\text{W}_{80}$.} \textbf{a}, \textbf{d} H diffusion coefficients at different temperatures using regular KMC (rKMC) and constrained KMC (cKMC), and VFT model fitting for $\text{Mo}_{25}\text{Nb}_{25}\text{Ta}_{25}\text{W}_{25}$ and $\text{Ta}_{20}\text{W}_{80}$, respectively. \textbf{b}, \textbf{e} Accessible diffusion barrier (DB) spectra at different temperatures and intrinsic DB for $\text{Mo}_{25}\text{Nb}_{25}\text{Ta}_{25}\text{W}_{25}$ and $\text{Ta}_{20}\text{W}_{80}$, respectively. \textbf{c}, \textbf{f} Accessible solution energy (SE) spectra at different temperatures and intrinsic SE for $\text{Mo}_{25}\text{Nb}_{25}\text{Ta}_{25}\text{W}_{25}$ and $\text{Ta}_{20}\text{W}_{80}$, respectively. The temperature-dependent evolution of the accessible DB and SE spectra corresponds to the diffusion curves produced by the cKMC method at the same temperatures.}
    \label{fig2}
\end{figure}

We first focus on the compositions $\text{Mo}_{25}\text{Nb}_{25}\text{Ta}_{25}\text{W}_{25}$ and $\text{Ta}_{20}\text{W}_{80}$, which represent equimolar and non-equimolar cases characterized by high and low mixing entropy levels, respectively. In Figs \ref{fig2}a and \ref{fig2}d, the diffusion curves obtained from cKMC are depicted as straight lines for both the compositions. This linear representation is indicative of the diffusion behavior solely influenced by temperature effects, which aligns well with the traditional Arrhenius equation. In contrast, the diffusion curves from rKMC are influenced by both the complex energy landscapes and temperature variations, resulting in a distinctly nonlinear trend, particularly at lower temperatures. This deviation from Arrhenius behavior towards super-Arrhenius behavior at low temperatures parallels the relaxation dynamics observed in glassy materials\cite{Hentschel2012,Hasyim2024}, where relaxation times increase disproportionately as temperatures decrease due to the emergence of cooperative atomic rearrangements and an increasingly complex energy landscape. Such super-Arrhenius behavior has also been captured in H diffusion in amorphous metals with deep trapping sites by simplified energy landscape\cite{KIRCHHEIM1988}. The Vogel-Fulcher-Tammann (VFT) model\cite{VFT} provides an appropriate framework to describe the nonlinear temperature dependencies in Figs \ref{fig2}a and \ref{fig2}d:
\begin{equation} \label{eq2}
    D = D_0 \exp\left(-\frac{Q_{\text{VFT}}}{k_B (T - T_0)}\right)
\end{equation}
where \( T_0 \) is the Vogel temperature, indicating a theoretical temperature below which diffusion ceases, and \( Q_{\text{VFT}} \) represents the effective activation energy. $D_0$ represents the pre-exponential factor, indicating the diffusion coefficient at infinite temperature. $k_{\text{B}}$ is the Boltzmann constant. Fitting the diffusion curves obtained from rKMC, we determine the values of $T_0$ as 62.11 K and 163.5 K for $\text{Mo}_{25}\text{Nb}_{25}\text{Ta}_{25}\text{W}_{25}$ and $\text{Ta}_{20}\text{W}_{80}$, respectively. This indicates a more pronounced non-linearity in the super-Arrhenius H diffusion behavior for the latter compared to the former. Moreover, the VFT model predicts $D = 4.09 \times 10^{-18} \, \text{m}^2/\text{s}$ for $\text{Ta}_{20}\text{W}_{80}$ at room temperature, which is significantly lower than $D = 1.44 \times 10^{-12} \, \text{m}^2/\text{s}$ for $\text{Mo}_{25}\text{Nb}_{25}\text{Ta}_{25}\text{W}_{25}$ at the same temperature. The discrepancy in six orders of magnitude highlights the considerable potential of non-equimolar compositions in H embrittlement mitigation by making H less diffusive. 

To investigate the mechanisms underlying such unique diffusion behavior, we analyze the accessible DB and SE spectra during rKMC simulations for the two compositions, as shown in Figs \ref{fig2}(b, c, e, f) alongside the intrinsic DB and SE spectra. It is observed that at low temperatures, H diffusion predominantly occurs by the jumps between T sites favoring paths with low SE and DB, as shown by a high probability of low-DB and low-SE regime. This indicates that low-SE T sites are related to the low-DB paths in random MPEAs. With increasing temperature, the accessible spectra converge towards the intrinsic spectra, corresponding to the convergence of straight lines from cKMC towards the curve from rKMC in Figs \ref{fig2}a and \ref{fig2}d.  This is due to the increased kinetic energy of H atom at higher temperatures, which enables it to overcome the high energy barriers more readily, thus facilitating faster diffusion within the MPEAs. The significant difference in the super-Arrhenius behaviors between $\text{Mo}_{25}\text{Nb}_{25}\text{Ta}_{25}\text{W}_{25}$ and $\text{Ta}_{20}\text{W}_{80}$ can be attributed to the disparities in their intrinsic DB and SE spectra. For $\text{Mo}_{25}\text{Nb}_{25}\text{Ta}_{25}\text{W}_{25}$, both the DB and SE exhibit a Gaussian-like distribution (Figs \ref{fig2}b and \ref{fig2}c), indicating that the majority of DBs and SEs concentrate around a central value, with fewer occurrences of extreme high or low values. Conversely, for $\text{Ta}_{20}\text{W}_{80}$, the distribution of intrinsic SE is positively skewed, displaying a long tail at the low SE end. This indicates a significant presence of extremely low-SE sites, which could act as deep trapping sites for H, thereby impeding its diffusion.

\begin{figure}[!ht]
    \centering
    \includegraphics[width=1\linewidth]{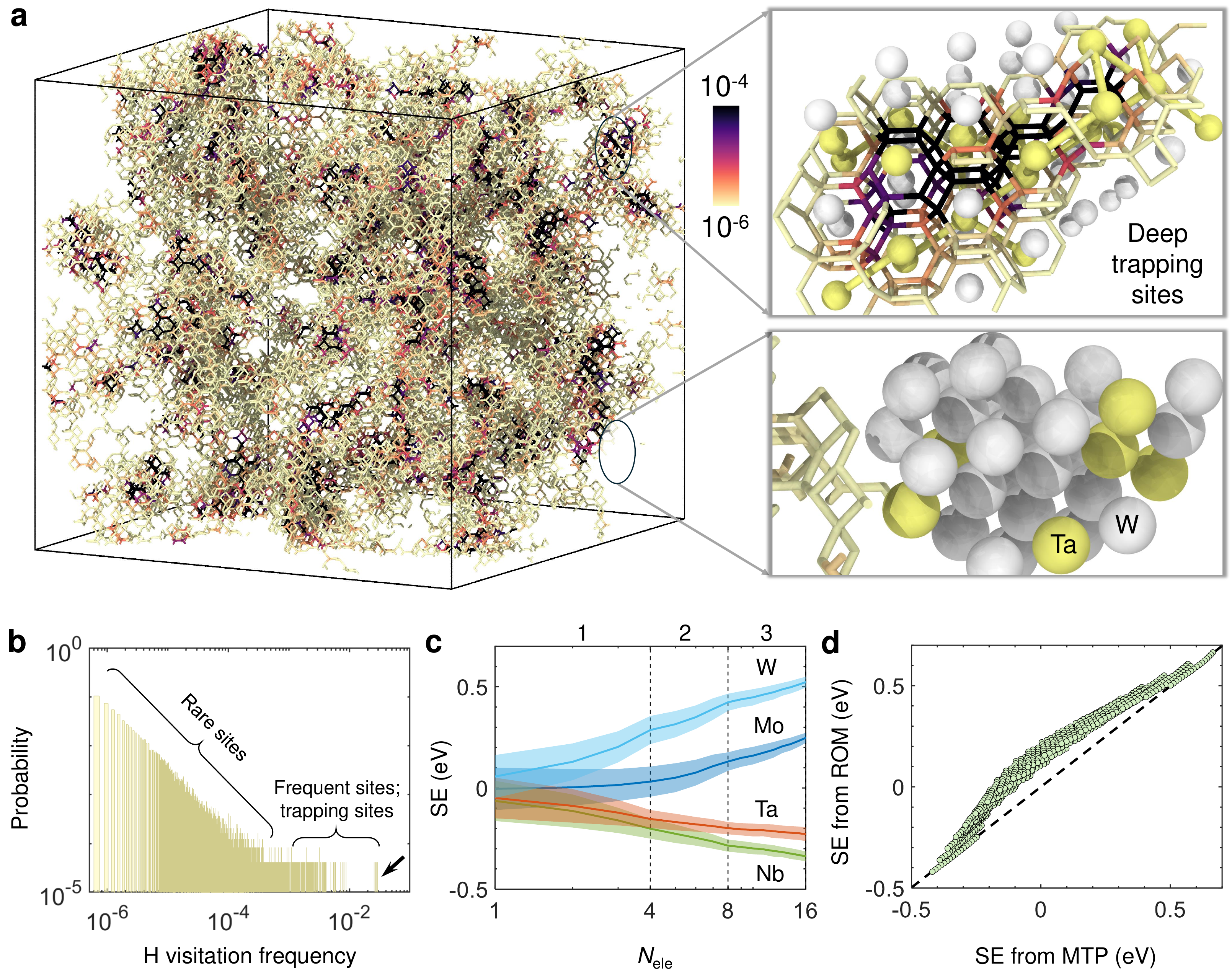}
    \caption{\textbf{Atomistic mechanisms of strong super-Arrhenius H diffusion in non-equimolar $\text{Ta}_{20}\text{W}_{80}$.} \textbf{a} Visualization of H visitation frequencies at all accessible tetrahedral (T) interstitial sites at 500 K in $\text{Ta}_{20}\text{W}_{80}$. The visitation frequency for each T site is calculated by dividing the visitation count of each site by the total number of jumps in the KMC simulations. The right top and bottom panels show the metallic environments near trapping sites and inaccessible sites, respectively. The H diffusion paths between two T sites are depicted as atomic bonds. \textbf{b} Probability distribution of visitation frequencies at 500 K for $\text{Ta}_{20}\text{W}_{80}$. \textbf{c} The general effects of metallic environments on H solution energy. The numbers 1-3 displayed at the top denote the number of shells corresponding to the nearest neighbors. \textbf{d} Prediction of H solution energy based on the rule of mixing (ROM).}
    \label{fig3}
\end{figure}

To further explore the atomistic mechanisms behind the pronounced super-Arrhenius behavior of H diffusion in $\text{Ta}_{20}\text{W}_{80}$, we present the H diffusion trajectories at 500 K, color-coded by visitation frequency, as depicted in Fig. \ref{fig3}a. The visitation frequency for each T site is calculated by dividing the visitation count of each site by the total number of jumps in the KMC simulations. Observations reveal that H is confined to specific regions of the $\text{Ta}_{20}\text{W}_{80}$ matrix, with most regions exhibiting very low visitation frequencies, while a few isolated regions show very high frequencies, indicating that H is predominantly trapped in these regions. Fig. \ref{fig3}b illustrates the distribution of visitation frequencies across all accessible T sites, spanning four orders of magnitude and heavily skewed towards lower frequencies ($10^{-6}$ to $10^{-4}$). Only a minimal number of T sites display very high frequencies ($10^{-4}$ to $10^{-2}$), acting as effective trapping sites. Consequently, H diffusion in $\text{Ta}_{20}\text{W}_{80}$ is highly non-uniform, attributable to the heterogeneous chemical environment. The right two panels of Fig. \ref{fig3}a highlight the metallic environments surrounding a deep trapping site (right top panel), and near a rarely visited site (right bottom panel). It is noted that a small cluster of Ta surrounds the deep trapping sites, while W-rich regions are areas that H cannot penetrate. This behavior can be rationalized by the distinct H solubility associated with different elements. Fig. \ref{fig3}c presents the SEs of H in various metallic environments, calculated by varying the number of each element surrounding the central H atom (see Fig. S10 for the calculation details and the results with different lattice constants). Nb and Ta exhibit negative SEs, indicating their affinity for H and categorizing them as “H-favoring” elements. Consequently, Nb- or Ta-rich regions have the potential to act as effective trapping sites for H. In contrast, Mo and W display positive SEs, indicating their role as “H-repelling” elements; thus, Mo- and W-rich regions inhibit H access. Furthermore, Fig. 3d illustrates that the rule of mixing can approximately predict the SE of MPEAs across 287 compositions based on the individual elemental SEs, suggesting that combinations of Nb or Ta with Mo or W result in a moderate SE. 

Considering the insights from Figs \ref{fig3}c and \ref{fig3}d, it becomes evident that isolated Nb/Ta atoms establish a metallic environment characterized by a favorable negative SE, while Mo/W-rich regions correspond to environments with very high SEs. Mixed Nb/Ta-Mo/W regions, on the other hand, exhibit moderate SEs with random values. This creates a roughened energy landscape for H diffusion in general MPEAs. The relatively low concentration of Ta in the $\text{Ta}_{20}\text{W}_{80}$ composition inevitably leads to the formation of isolated small Ta clusters surrounded by W atoms, which consequently results in the prominent long tail at the low SE end observed in Fig. \ref{fig2}f. These isolated small Ta clusters act as strong trapping sites for H (Fig. \ref{fig3}a), significantly suppressing its diffusion. Conversely, $\text{Mo}_{25}\text{Nb}_{25}\text{Ta}_{25}\text{W}_{25}$ contains equal concentrations of “H-favoring” elements (Nb and Ta) and “H-repelling” elements (Mo and W). Random mixing creates a relatively uniform energy landscape because each Nb or Ta neighbor is likely accompanied by a Mo or W neighbor, preventing the formation of very deep trapping sites and leading to the weak super-Arrhenius behavior.

\subsection*{Compositional space study and symbolic machine learning regression}

\begin{figure}[!ht]
    \centering
    \includegraphics[width=1\linewidth]{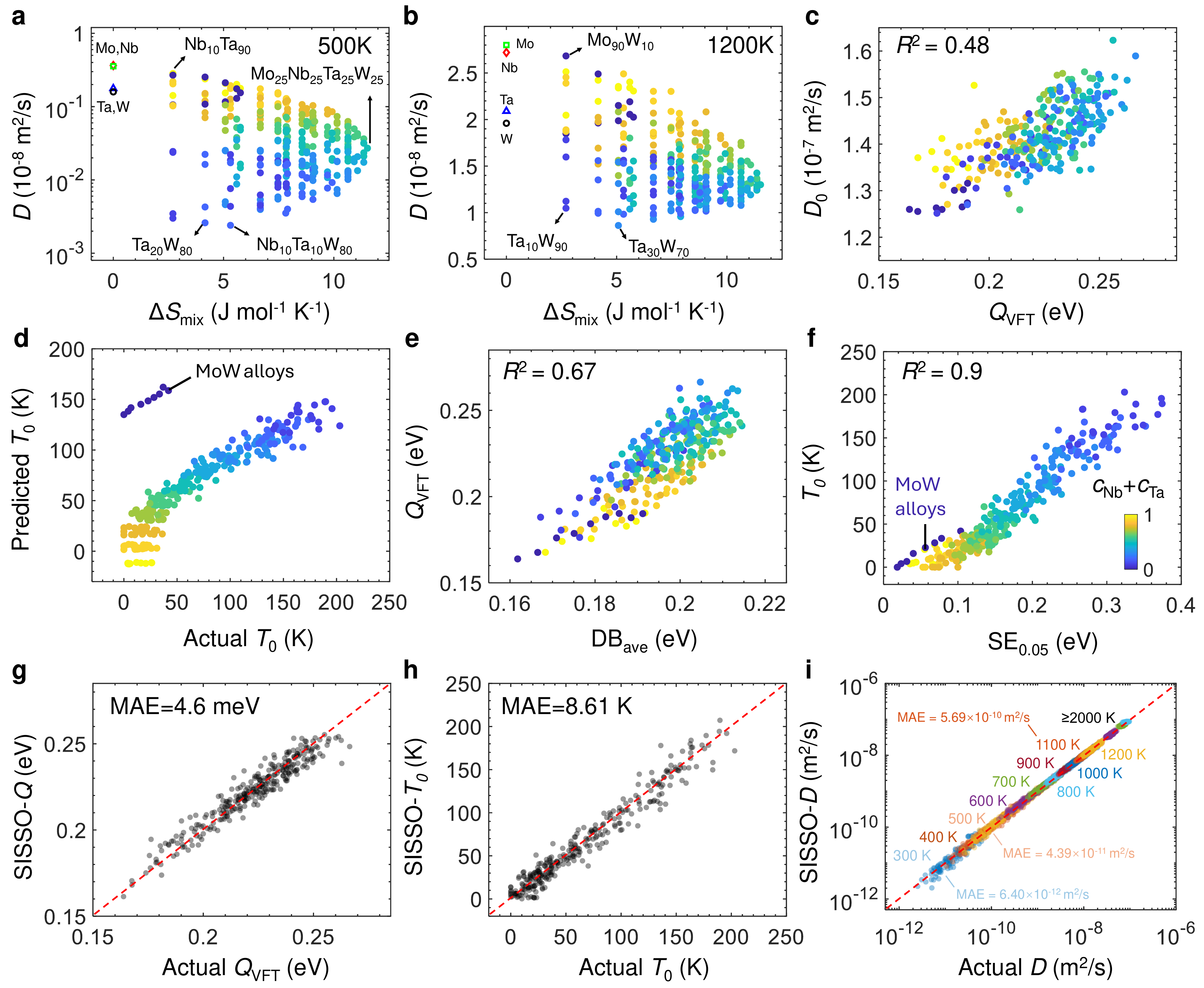}
    \caption{\textbf{H diffusion dynamics across the entire compositional spectrum in MoNbTaW systems, encompassing 287 compositions.} \textbf{a}, \textbf{b} Diffusion coefficients $D$ vs. mixing entropy $S_{\textbf{mix}}$ for 287 compositions at 500 K and 1200 K. \textbf{c} Correlation between the prefactor $D_0$ and the effective activation energy $Q_\text{VFT}$. \textbf{d} Predicted $T_0$ from Eq. \ref{eq4} in the main text versus actual $T_0$ determined by KMC simulations.\textbf{e} Correlations between $Q_{\text{VFT}}$ and the average of the intrinsic diffusion barrier ($\text{DB}_{\text{ave}}$). \textbf{f} Correlation between the Vogel temperature $T_0$ and the lower 5\% percentile of solution energy ($\text{SE}_{\text{0.05}}$). All data points in figures a-c are color-coded by the combined concentrations of Nb and Ta ($c_{\text{Nb}}+c_{\text{Ta}}$). The results depicted in figures \textbf{a}-\textbf{c} demonstrate that $c_{\text{Nb}}+c_{\text{Ta}}$ is a crucial factor influencing H diffusion. \textbf{g}-\textbf{i} Machine-learning symbolic regression (MLSR) for predicting $Q_{\text{VFT}}$, $T_0$ and H diffusion coefficients in the full compositional space across a wide temperature range from 300 K to 5000 K.}
    \label{fig4}
\end{figure}

In the investigation of H diffusion dynamics across 287 compositions, we first examine the diffusion coefficients at 500 K for all compositions, alongside four pure elements, as depicted in Fig. \ref{fig4}a. At this relatively low temperature, we observe that most compositions exhibit lower diffusivity compared to the pure metals, indicative of a pronounced sluggish diffusion effect. Notably, $\text{Mo}_{25}\text{Nb}_{25}\text{Ta}_{25}\text{W}_{25}$ demonstrates moderate diffusivity, while non-equimolar compositions display extremes in diffusivity. For instance, $\text{Nb}_{10}\text{Ta}_{90}$ shows diffusivity surpassing that of pure Ta and W, suggesting that the inclusion of a minor amount of Nb enhances the H diffusivity in Ta. In contrast, compositions such as $\text{Ta}_{20}\text{W}_{80}$ and $\text{Nb}_{10}\text{Ta}_{10}\text{W}_{80}$ exhibit markedly low diffusivity, with the diffusivity in $\text{Nb}_{10}\text{Ta}_{90}$ being 118 times higher than that of $\text{Nb}_{10}\text{Ta}_{10}\text{W}_{80}$. At the elevated temperature of 1200 K, as shown in Fig. \ref{fig4}b, the diffusivity of all compositions falls within the range of $8.62 \times 10^{-9}$ to $2.68 \times 10^{-8} \, \text{m}^2/\text{s}$, aligning closely with the range observed for pure metals, which is between $1.96 \times 10^{-8}$ to $2.8 \times 10^{-8} \, \text{m}^2/\text{s}$. This observation suggests that the sluggish diffusion effect is substantially diminished at high temperatures. 

We next utilize the VFT model to fit the diffusion coefficients of 287 compositions, from which we extract the key parameters: the pre-exponential factor ($D_0$), the effective activation energy ($Q_{\text{VFT}}$), and the Vogel temperature ($T_0$). One notable finding is the discernible correlation between $D_0$ and $Q_{\text{VFT}}$, as illustrated in Fig. \ref{fig4}c. This relationship can be understood through the lens of the  Meyer-Neldel rule\cite{Yelon1992}. Based on this correlation, we proceed to fit a linear expression for the logarithm of $D_0$ in terms of $Q_{\text{VFT}}$:
\begin{equation} \label{eq3}
    \log(D_0) = 1.634 Q_{\text{VFT}} - 16.13
\end{equation}
which can be used to predict $D_0$ based on $Q_{\text{VFT}}$. Moreover, the narrow range of $D_0$ distribution across 287 compositions suggests comparable H diffusivity at higher temperatures among different MPEAs. For the Vogel temperature $T_0$, we also fit a concentration-dependent linear expression:
\begin{equation} \label{eq4}
    T_0 = 130.00\,c_{\text{Mo}} - 10.08\,c_{\text{Nb}} - 10.82\,c_{\text{Ta}} + 175.43\,c_{\text{W}}
\end{equation}
where $c$ is the element concentration. The correlation between predicted and the actual values of $T_0$ fitted by KMC results is depicted in Fig. \ref{fig4}d. Our analysis demonstrates that, aside from MoW binary alloys, \( T_0 \) adheres closely to a concentration-dependent linear model. Notably, \( T_0 \) exhibits a general decline as the cumulative concentrations of Nb and Ta (\( c_{\text{Nb}} + c_{\text{Ta}} \)) increase. Further exploration into the correlations between $Q_{\text{VFT}}$ and $T_0$ with different statistical measurements of DB and SE spectra, has yielded intriguing results. Specifically, we find that $Q_{\text{VFT}}$ is positively correlated with the average value of intrinsic DB ($\text{DB}_{\text{ave}}$), and $T_0$ is positively correlated with the 5th percentile of SE ($\text{SE}_{0.05}$), as illustrated in Figs \ref{fig4}e and \ref{fig4}f, respectively. Another interesting observation is that \( c_{\text{Nb}} + c_{\text{Ta}} \) plays a dominant role in H diffusion. This is indicated by the varying marker colors in Figs \ref{fig4}a-\ref{fig4}f. To enhance the accuracy of our models for predicting $Q_{\text{VFT}}$ and $T_0$, we have employed an advanced machine-learning symbolic regression (MLSR) technique, known as the sure independence screening and sparsifying operator (SISSO) method\cite{Ouyang2018}, to perform symbolic regression. The details of data preparation and SISSO training process are provided in “Methods”. The 5-fold and 10-fold cross-validation results shown in the Supplementary Note 4 demonstrate that the SISSO method does not suffer from overfitting in predicting $Q_{\text{VFT}}$ and $T_0$. Therefore, we employ all available data to derive analytical expressions based on the statistical properties of intrinsic DB and SE spectra:
\begin{equation} \label{eq5}
    Q_{\text{VFT}} (\text{eV}) = 0.4875 \left[ \left( 2 \text{DB}_{\text{ave}} + \text{DB}_{\text{IQR}} - \text{DB}_{\text{sd}} \right) + \left( \text{SE}_{\text{sd}} + \text{SE}_{\text{IQR}} \right)\frac{\text{SE}_{\text{IQR}}}{\text{SE}_{\text{range}}} \right]
\end{equation}
\begin{equation} \label{eq6}
    T_0 (\text{K}) = 834.46 \left( \text{SE}_{0.05} - 2\text{SE}_{\text{sd}} + \text{DB}_{\text{sd}} \right) \left| \frac{\text{DB}_{\text{sd}}}{\text{DB}_{\text{ave}}} - \frac{\text{SE}_{\text{sk}}}{\text{SE}_{\text{kt}}} \right|
\end{equation}
where $\text{DB}_{\text{ave}}$, $\text{DB}_{\text{IQR}}$, and $\text{DB}_{\text{sd}}$ represent the average, interquartile range, and standard deviation of DB spectrum, respectively and $\text{SE}_{\text{sd}}$, $\text{SE}_{\text{IQR}}$, $\text{SE}_{\text{range}}$, $\text{SE}_{\text{0.05}}$, $\text{SE}_{\text{sk}}$, $\text{SE}_{\text{kt}}$ are the standard deviation, interquartile range, range, 5th percentile, skewness and kurtosis of SE, respectively. Eq. \ref{eq5} shows that the effective activation energy ($Q_{\text{VFT}}$) is primarily determined by the average of the DB spectrum and is influenced by the statistical dispersion of DB. Additionally, the dispersion of the SE also plays a role. On the other hand, Eq. \ref{eq6} indicates that the Vogel temperature ($T_0$) is primarily influenced by $\text{SE}_{\text{0.05}}$. Both $\text{SE}_{\text{0.05}}$ and $\text{DB}_{\text{sd}}$ show a positive correlation with $T_0$, whereas the $\text{SE}_{\text{sd}}$ is negatively correlated with $T_0$. Additionally, $T_0$ is influenced by the relative dispersion of DB and the distribution shape of SE. One of the most important findings of this work is the pivotal role of $\text{SE}_{\text{0.05}}$ in determining the Vogel temperature ($T_0$), as evidenced by Fig. \ref{fig4}f and Eq. \ref{eq6}. Physically, $\text{SE}_{\text{0.05}}$ reflects the relative strength of trapping H in a system, with a high $\text{SE}_{\text{0.05}}$ indicating a small number of deep trapping sites. Given that $T_0$ is a key parameter influencing the non-linear behavior of super-Arrhenius diffusion, a high $T_0$ is crucial to potentially suppress or even halt H diffusion at room temperature. This finding is particularly significant for mitigating H embrittlement, as MPEAs hold the potential to effectively prevent H diffusion at room temperature. This could be the underlying reason for the experimentally reported exceptional H embrittlement resistance in FCC MPEAs such as CrCoFeMnNi, CrCoNi and CoNiV\cite{Luo2017,Luo2018-1,Luo2018-2,Luo2020}.

Figs \ref{fig4}g and \ref{fig4}h illustrate that our SISSO models proficiently predict the parameters $Q_{\text{VFT}}$ and $T_0$ across the entire compositions. By combining Eqs \ref{eq3}, \ref{eq5}, and \ref{eq6}, we can accurately predict diffusion coefficients for BCC MPEAs at any temperature, using only the statistical features of DB and SE spectra as inputs. The effectiveness of our model is demonstrated in Fig. \ref{fig4}i, where we compare diffusion coefficients from KMC and SISSO models across 287 compositions at various temperatures. The low prediction errors underscore the accuracy of our SISSO models. It is important to highlight that the model, derived from SISSO-generated expressions, is analytical and exhibits high transferability, enabling it to predict H diffusivity in other BCC MPEAs at diverse temperatures.

\subsection*{Chemical short-range order effect on the super-Arrhenius behavior}
\begin{figure}[H]
    \centering
    \includegraphics[width=0.8\linewidth]{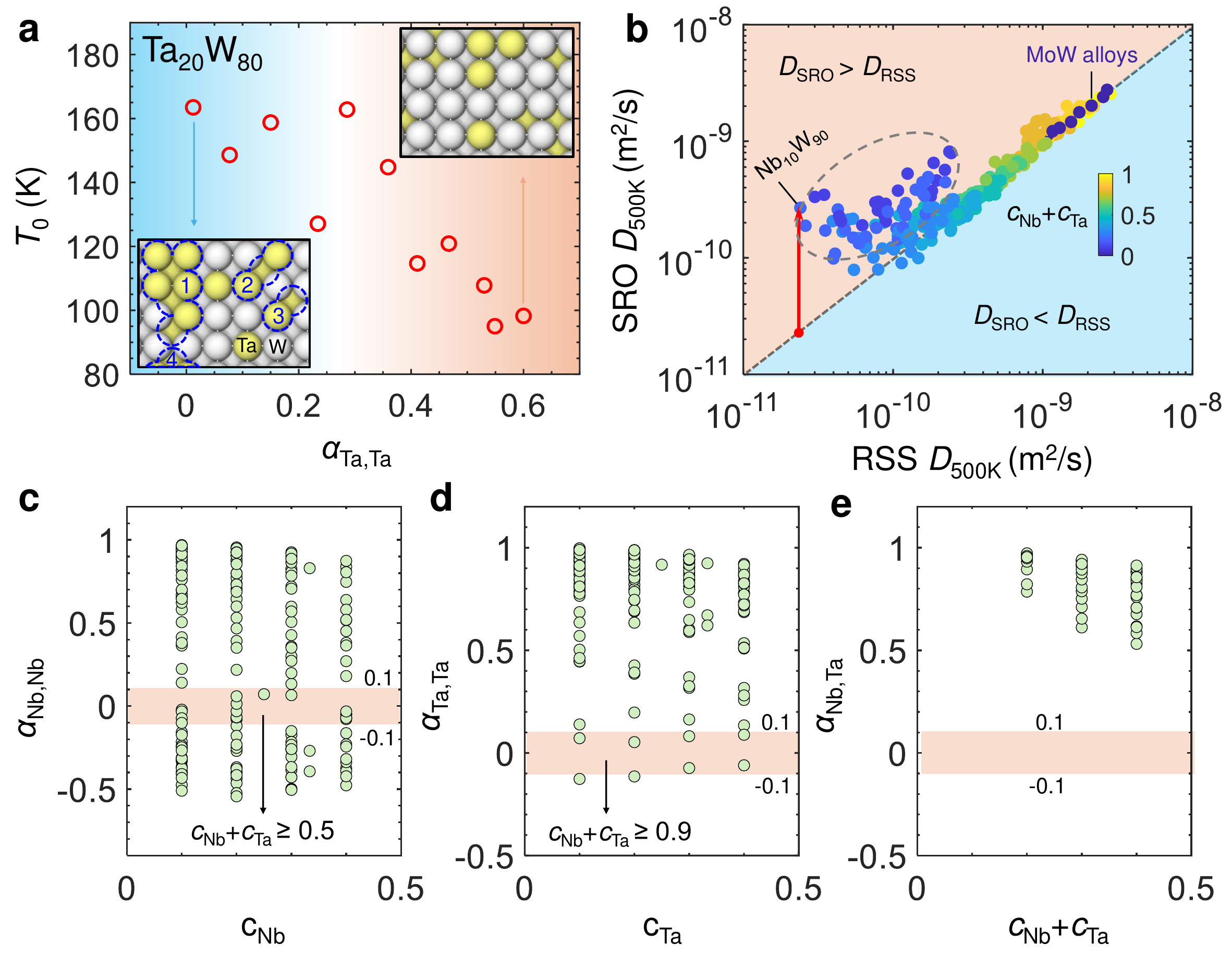}
    \caption{\textbf{The effect of chemical short-rang order on the H super-Arrhenius diffusion behaviors in MoNbTaW systems.} \textbf{a} Variation of the Vogel temperature ($T_0$) with respect to the WC parameter of Nb-Nb (\(\alpha_{\text{Ta,Ta}}\), \(\alpha_{\text{Ta,Ta}}\), and \(\alpha_{\text{Nb,Ta}}\)). The left inset shows the small Ta cluster in random solid solution (RSS) of $\text{Ta}_{20}\text{W}_{80}$, and the right inset shows the dispersed Ta atoms in $\text{Ta}_{20}\text{W}_{80}$ with SRO. \textbf{b} Comparation of H diffusion coefficients at 500 K for RSS and SRO across 287 compositions. \textbf{c}-\textbf{e} Distribution of \(\alpha_{\text{Nb,Nb}}\), \(\alpha_{\text{Ta,Ta}}\), and \(\alpha_{\text{Nb,Ta}}\) in MoNbTaW alloys with respect to Nb concentration ($c_{\text{Nb}}$), Ta concentration ($c_{\text{Nb}}$) and the combined concentration of Nb and Ta ($c_{\text{Nb}}+c_{\text{Ta}}$), respectively. The shaded regions in \textbf{c}-\textbf{e} represent areas of weak SRO, where the WC parameters satisfy the condition $|\alpha| < 0.1$.}
    \label{fig5}
\end{figure}

Chemical short-range order (SRO) is a ubiquitous phenomenon in MPEAs\cite{Han2024} and significantly influences various properties such as strength\cite{Zhang2023}, ductility\cite{Chen2021}, and vacancy diffusion\cite{Xing2024}. This highlights the critical need to examine its effects on H diffusion. Fig. \ref{fig3} underscores the essential role of small Ta clusters in modulating H diffusivity within the $\text{Ta}_{20}\text{W}_{80}$ alloy. Given the pronounced impact of SRO on elemental distribution, it is anticipated that the formation of SRO markedly influences the super-Arrhenius diffusion dynamics of H in MPEAs. We employ our NN-KMC approach to compute H diffusivity across 287 alloy compositions, using configurations derived from hybrid Monte Carlo/molecular dynamics (MC/MD) simulations (see "Methods"), which facilitate the emergence of SRO. As a specific case study, we present the Vogel temperatures ($T_0$)  fitted by KMC simulations corresponding to the various degrees of SRO in $\text{Ta}_{20}\text{W}_{80}$, measured by the Warren-Cowley (WC) parameter of Ta atoms (\(\alpha_{\text{Ta,Ta}}\)), as shown in Fig. \ref{fig5}a. It is observed that $T_0$ decreases with increasing \(\alpha_{\text{Ta,Ta}}\), indicating that SRO significantly suppresses super-Arrhenius dynamics. The left inset Fig. \ref{fig5}a illustrates these isolated small Ta clusters (identified as 1-4 with blue dashed lines) within a random solid solution (RSS) of the $\text{Ta}_{20}\text{W}_{80}$ alloy with \(\alpha_{\text{Ta,Ta}}\) = 0, serving as effective trapping sites for H, based on the observations in Fig. \ref{fig3}. Post-hybrid MC/MD, the Ta atoms are depicted as being sparsely distributed throughout the $\text{Ta}_{20}\text{W}_{80}$ alloy, as demonstrated in the right inset of Fig. \ref{fig5}a. The WC parameters, \(\alpha_{\text{Ta,Ta}}\) = 0.60 and \(\alpha_{\text{Ta,W}}\) = -0.15, reflect the strong repulsion between Ta atoms and the attraction between Ta and W atoms. This altered atomic arrangement leads to the formation of a local B2 structure with Ta-W pairs (the left inset in Fig. \ref{fig5}a), increasing H solution energies near the Ta atoms, which diminishes the effectiveness of the deep trapping sites formed by small Ta clusters, thereby contributing to the observed reduction in $T_0$.

To elucidate the general effects of SRO on H diffusion, we compute the diffusion coefficients for 287 alloy compositions at 500 K with established SRO in MC/MD simulations, comparing them to their RSS counterparts, as shown in Fig. \ref{fig5}b. Additionally, a detailed analysis of the SRO effect across all equimolar compositions is presented in Fig. S11. Interestingly, while SRO does not significantly affect H diffusivity in compositions with high coefficients in RSS, it does enhance diffusivity in compositions with initially low diffusion coefficients as indicated by the dashed circle in Fig. \ref{fig5}b. One notable composition is $\text{Nb}_{10}\text{W}_{90}$, which exhibits a significant increase in the diffusion coefficient from $2.40 \times 10^{-11} \, \text{m}^2/\text{s}$ to $2.68 \times 10^{-10} \, \text{m}^2/\text{s}$ after introducing SRO, as indicated by the red arrow in Fig. \ref{fig5}b. This differential impact of SRO on H diffusivity correlates strongly with $c_{\text{Nb}}+c_{\text{Ta}}$. Specifically, significant effects of SRO are observed predominantly in compositions with low $c_{\text{Nb}}+c_{\text{Ta}}$, as illustrated by the varying marker colors in Fig. \ref{fig5}b. This pattern suggests that the interactions between H and the metal elements, as well as the interactions among the metal elements themselves that contribute to SRO formation, play critical roles. As previously demonstrated in Fig. \ref{fig4}, alloys with low $c_{\text{Nb}}+c_{\text{Ta}}$ typically exhibit reduced H diffusivity. This reduction is attributed to the H-favoring characteristics of Nb and Ta, wherein isolated atomic clusters of these elements function as potent trapping sites, thereby impeding H diffusion. The formation of SRO modifies the distribution and presence of these clusters, influencing H mobility. In exploring how SRO influences elemental distribution, we focused on alloys where $c_{\text{Nb}}$, $c_{\text{Ta}}$, and their combined concentrations do not exceed 0.4. The corresponding WC parameters for Nb-Nb, Ta-Ta, and Nb-Ta interactions (\(\alpha_{\text{Nb,Nb}}\), \(\alpha_{\text{Ta,Ta}}\), and \(\alpha_{\text{Nb,Ta}}\)) are detailed in Figs \ref{fig5}c-e. Analysis of Nb-Nb interactions (Fig. \ref{fig5}c) reveals a dominance of either strong attraction or repulsion for low $c_{\text{Nb}}+c_{\text{Ta}}$. Strong repulsion leads to the dispersion of Nb atoms and disruption of small clusters, analogous to the behavior of Ta as observed in the left inset of Fig. \ref{fig5}a. On the other hand, strong attraction results in the formation of connected Nb regions, potentially creating fast H diffusion pathways. Fig. S12f provides an example of this phenomenon in an equimolar NbTaW alloy, where Nb atoms segregate into a distinct layer and as a result, H diffusion in NbTaW degenerates into H diffusion within the pure Nb layer. In the case of Ta-Ta interactions shown in Fig. \ref{fig5}d, strong repulsion is predominantly observed, whereas weak interactions mainly occur in compositions with a high $c_{\text{Nb}} + c_{\text{Ta}}$. Nb-Ta interactions consistently exhibit strong repulsion across all examined compositions (Fig. \ref{fig5}e). Thus, the formation of SRO in low $c_{\text{Nb}}+c_{\text{Ta}}$ compositions, characterized by strong elemental interactions, disrupts the deep trapping sites typically present in RSS, facilitating H diffusion.

\section*{Discussion}

Exploring the complex mechanisms of H diffusion in MPEAs is crucial not only for advancing our fundamental understanding of dynamics in disordered systems but also for driving innovations in materials engineering that address critical societal needs\cite{Cheng2023}. Insights into H interactions within these alloys could revolutionize industries by developing more durable, efficient, and environmentally friendly materials, thus offering significant benefits in fields ranging from renewable energy to nuclear fusion engineering\cite{Odenweller2022,El-Atwani-2023}. Traditional experimental techniques often prove cost-prohibitive and lack the necessary resolution for effective detection of H within materials\cite{Fukai1985}. Computationally, the investigation of H diffusion in MPEAs presents a formidable challenge due to its inherent multiscale characteristics across both spatial and temporal dimensions. In terms of time scales, our findings (illustrated in Fig. \ref{fig3}) reveal a clear difference in the dynamics of H diffusion: rapid within trapping sites yet markedly slow between them. This contrasting diffusion behavior underscores the limitation of studying H diffusion at lower temperatures using MD, where the slower jumps between deep trapping sites are particularly difficult to capture accurately due to the limited timescale in MD. Spatially, H diffusion represents a multi-length scale phenomenon. As evidenced by our analyses in Fig. S7, employing a smaller supercell containing fewer atoms leads to significant fluctuations in the DB spectrum and correspondingly, the H diffusion behavior. This variability can be intuitively understood through statistical principles, where the chemical heterogeneity of a random MPEA necessitates a substantial atoms to correctly represent macroscopic diffusion behaviors. Collectively, these multiscale attributes of H diffusion in MPEAs pose substantial challenges to conventional simulation methodologies such as MD and KMC, highlighting the need for innovative approaches in accurately capturing these complex diffusion dynamics. 

To address these multiscale challenges, this work introduces a multi-faceted machine learning computational framework designed to effectively capture the complex dynamics of H diffusion in MPEAs. The framework comprises several innovative components: machine-learning force fields (MLFF), neural network-driven Kinetic Monte Carlo (NN-KMC), and machine-learning symbolic regression (MLSR). Each component is tailored to tackle specific facets of the H diffusion process, enabling accurate simulation and analysis. MLFF is utilized to generate accurate DB database within a diverse alloy composition, significantly enhancing the simulation fidelity with DFT accuracy. NN-KMC is applied to efficiently process and predict the stochastic behavior of H atoms as they navigate through the complex energy landscape of MPEAs, thus speeding up the kinetic simulations without sacrificing accuracy. Meanwhile, MLSR techniques help in extracting meaningful physical insights from complex data patterns, facilitating a deeper understanding of the underlying diffusion mechanisms. This significant advancement paves the way for investigations into H diffusion dynamics across various BCC and FCC MPEA systems, given the transferability of our method to all classes of MPEAs. Although our developed MLFF currently only considers four BCC refractory elements (Mo, Nb, Ta, and W) and their interactions, it serves as a valuable tool for probing H diffusion mechanisms in complex systems with rugged energy landscapes. The analytical expressions obtained by MLSR, along with the VFT model, provide a complete model for predicting H diffusion in BCC MPEAs in general. Future work will expand our framework to incorporate the effects of defects such as vacancies, dislocations, and grain boundaries, as well as the influence of varying H concentrations.

The DPS strategy introduced in this study demonstrates significant improvements over previous active learning strategies for the MLFF development. Traditionally, active learning strategies progressively learn unpredictable configurations either from ab initio MD or from MLFF-based MD simulations by an on-the-fly manner\cite{Jinnouchi2020,Novikov2021}. However, our KMC results presented in Fig. \ref{fig3} suggest that H diffusion in MPEAs can be extremely slow, and many environments remain inaccessible even after prolonged periods. Therefore, the extended period required for diffusion at low temperatures presents a primary challenge. Consequently, the resultant MLFF cannot be applied to arbitrary compositions of MPEAs, making it impossible to accurately determine the intrinsic DB and SE spectra. The proposed DPS strategy addresses these issues by leveraging the fact that a single H atom does not alter the lattice structure of MPEA matrix. This insight allows us to circumvent the need for computationally expensive active learning-based ab initio MD trajectories or on-the-fly methods. Instead, DPS efficiently and effectively captures all representative configurations across diverse metallic environments by accounting for lattice distortion and thermal vibration at different temperatures. This approach results in a highly accurate and comprehensive MLFF, as demonstrated by our extensive validation results shown in Fig. S2. 

It is important to note that the development of a new MLFF for the MoNbTaW-H system plays a crucial role in this study. Although universal machine learning potentials such as CHGNET\cite{Deng2023}, and MACE\cite{Batatia2023} have been developed to model interactions across the periodic table, they often serve merely as foundation models and exhibit high errors; fine-tuned training is essential for specific studies. We have evaluated the performance of the latest MACE foundation model (mace-mp-0) against our MLFF in predicting the energy of MoNbTaW-H systems, as illustrated in Fig. S12. The results demonstrate that our MLFF offers significantly higher accuracy and efficiency than MACE. Additionally, we utilize KMC simulations to study H diffusion, employing energy barriers at 0 K to estimate event rates. To assess the thermal effects on diffusion dynamics, we compare the diffusion coefficients obtained from direct MD simulations and KMC simulations across 969 compositions using a small 4 × 4 × 4 supercell in Fig. S13. The consistent results indicate that the thermal effects will not change the relative diffusivity of different compositions.  Importantly, our new MLFF is also capable of investigating nuclear quantum effects at low temperatures using long-time path integral simulations such as path integral molecular dynamics (PIMD) for MPEAs\cite{Kwon2023}, although this application is not explored in the current study. We anticipate that nuclear quantum effects may mitigate the super-Arrhenius behavior. This aspect could be explored in future work.

Our study offers profound insights into H diffusivity across a broad spectrum of MoNbTaW MPEAs. Our analysis from an energy landscape perspective indicates that the randomness of DB is not the primary determinant for H diffusion in these alloys. Instead, the randomness of SE plays a pivotal role. This is because, unlike DB, which only accounts for individual jump events, the SE distribution offers a comprehensive view of the entire energy landscape, thereby influencing H diffusivity on a macroscopic scale. Focusing on metallic environments, our findings underscore that the presence of “H-favoring” elements significantly governs H diffusivity in random MPEAs and affects SRO effects in ordered MPEAs. Particularly, we spotlight Nb and Ta, the primary elements discussed in our work, as critical to understanding these dynamics. Additionally, other BCC transition metals like V, Ti, Zr, and Hf, known for their low SE\cite{Brouwer1989}, also play a crucial role. Introducing these elements, especially in low concentrations, alongside “H-repelling” elements such as W and Mo can lead to the formation of deep trapping sites. This interaction culminates in marked super-Arrhenius diffusion behaviors, which are essential for tailoring the H diffusivity properties of next-generation MPEAs for H-related applications. Regarding the SRO effect, our findings suggest that SRO typically has no impact on H diffusion. It facilitates H diffusion only when “H-favoring” elements are present in low concentrations. Therefore, careful selection of elements an is crucial to mitigate the potential effects of SRO.

Lastly, our study highlights the significant potential of MPEAs in mitigating H embrittlement, a long-standing issue in traditional alloys like aluminum alloys and steel\cite{Dwivedi2018}. In these conventional alloys, H rapidly diffuses through the lattice, accumulating at defects such as vacancies and grain boundaries that act as deep trapping sites, leading to material degradation and failure\cite{Chen2024}. In contrast, our findings suggest a different mechanism in MPEAs. As demonstrated in Figs \ref{fig2} and \ref{fig3}, the lattice of MPEAs, characterized by random chemical environments, intrinsically forms deep trapping sites. These sites, created by small clusters of “H-favoring” elements, are not only comparable but potentially more effective in trapping H than the conventional defects in MPEAs\cite{Yang2024}, providing enhanced resistance to H embrittlement. Traditional alloys often incorporate second phases or nano-precipitates to trap H, which slows its diffusion and prevents accumulation at critical sites\cite{Zhao2022,Sun2021}. MPEAs, however, benefit from a diverse arrangement of elements that intrinsically create a complex chemical environment, which enables the distribution of the trapping sites more uniformly throughout the alloys, eliminating the need for additional phases to trap H. This inherent feature of MPEAs promises a more robust and effective mitigation of H embrittlement. Furthermore, the specific arrangement and concentration of elements in MPEAs can be tailored to optimize other properties such as mechanical strength, ductility and radiation resistance, offering a customizable approach for various industrial applications. 

To summarize, we develop a comprehensive machine-learning computational framework to reveal the hidden dynamics of super-Arrhenius H diffusion in MPEAs. Our comprehensive high-throughput screening has unveiled the fundamental mechanisms governing super-Arrhenius H diffusion in random MPEAs and the effect of SRO, along with shedding light on how metallic environments affect H trapping. These findings emphasize the vital role of machine learning methodologies and their profound impact on steering the development of new materials to address the challenges in H-related applications.
\section*{Methods}
\subsection*{D-optimality-based pre-selection}
We employ the D-optimality-based Pre-Selection (DPS) to select representative H-MPEA configurations for machine-learning force field development\cite{Novikov2021}. In the framework of moment tensor-based MLFF, the energy of a configuration can be expressed as:
\begin{equation} \label{eq9}
E^{\text{mtp}}(\text{cfg}; \xi) = \sum_i \sum_{\alpha=1}^m \xi_\alpha B_\alpha(n_i) = \sum_{\alpha=1}^m \xi_\alpha \underbrace{\left(\sum_i B_\alpha(n_i)\right)}_{b_\alpha(\text{cfg})}
\end{equation}
where $n_i$ is the atomic environment of atom $i$,  $\xi_{\alpha}$ are the parameters to be fitted, \( B_{\alpha} \) are the basis functions, cfg is the abbreviation of configuration. When fitting to the energy values, an overdetermined system of K linear equations on $\xi$ is needed to solve with the matrix:
\begin{equation} \label{eq10}
B = \begin{bmatrix}
b_1(\text{cfg}_1) & \cdots & b_m(\text{cfg}_1) \\
\vdots & \ddots & \vdots \\
b_1(\text{cfg}_K) & \cdots & b_m(\text{cfg}_K)
\end{bmatrix}
\end{equation}
Following the D-optimality criterion, we strategically choose $m$ configurations to generate the most linearly independent equations, ensuring that the corresponding $m \times m$ submatrix $\textbf{A}$ achieves the maximum modulus of determinant, denoted as $|\det(\textbf{A})|$. The $m$ selected configurations are called as the representative configurations which corresponds to the most extreme and diverse ones. The DPS operations are conducted by the MLIP-2 package\cite{Novikov2021}.

\subsection*{DFT calculations}
We perform DFT calculations using the Vienna Ab initio Simulation Package (VASP)\cite{Kresse1996} with the Perdew--Burke--Ernzerhof (PBE)\cite{Perdew1996} exchange-correlation functional and projector-augmented plane wave\cite{PAW} potentials with a plane-wave cutoff energy of 520 eV. A consistent K-point density of $0.03 \times 2\pi/\text{\AA}$ using the Monkhorst-Pack Scheme is maintained using the Python tool VASPKIT\cite{Wang2021}. The energy threshold for self-consistency and the force threshold for structure relaxation are $10^{-6}$ eV and $0.01$ eV/\AA, respectively.

\subsection*{Machine-learning force field training database generation}
Fig. S1 shows the schematic of our database generation for the machine-learning force field. Our starting point is the established DFT database for BCC MoNbTaW MPEAs\cite{Li2020,Yin2021}. We intentionally exclude configurations with exposed surfaces, as our study concentrates exclusively on investigating H diffusion within defect-free bulk systems. On investigating the interactions of H with individual unary systems (Mo, Nb, Ta, and W), we leverage configurations derived from recent studies probing H-W and H-Nb interactions\cite{Kwon2023}. These configurations are adjusted through rescaling lattice constant to acquire configurations specific to H-Mo and H-Ta. The database comprises of H configurations at stable tetrahedral (T) sites, minimum energy paths obtained using the climbing-image Nudged Elastic Band (CI-NEB) under various strain levels for H diffusion, and non-equilibrium configurations obtained by active learning throughout classical MD and path integral molecular dynamics (PIMD) simulations\cite{Kwon2023}. This comprehensive dataset guarantees reliable simulations of H diffusion within elemental systems across a diverse temperature range. In the context of H-binary systems (MoNb, MoTa, MoW, NbTa, NbW, and TaW), we consider scenarios where H occupies all T sites, Octahedral (O) sites, and intermediary positions between T sites within a 36-atom equimolar special quasi-random structure (SQS)\cite{SQS}. We specifically select configurations from the initial four ionic relaxation steps in the conjugate gradient relaxation for each scenario. In exploring H-ternary systems (MoNbTa, MoNbW, MoTaW, NbTaW) and the H-quaternary system (MoNbTaW), we consider the configurations involving the full relaxation process, encompassing scenarios where H occupies all T sites within a 36-atom equimolar SQS. More importantly, we select representative configurations using DPS at 300 K, 700 K and 1200 K for all equimolar ternary and quaternary systems. A total of 85,762 configurations are used for training, with an additional 9,530 configurations allocated for testing. Moreover, the validation task in Fig. S2 involves an extensive dataset of over 40,000 configurations. Altogether, our database comprises approximately 140,000 configurations.

\subsection*{Machine-learning force field framework}
The moment tensor-based MLFF\cite{Novikov2021} and its formalism have been extensively studied and applied in the MPEAs\cite{Yin2021} and H-metal systems\cite{Kwon2023}, which essentially constructs contracted rotationally invariant local environment descriptors for each atom in the system and builds a polynomial regressed correlation between the potential energy surface (PES) and these descriptors. The descriptors, named moment tensors, are devised as follows:
\begin{equation} \label{eq11}
M_{\mu,\nu}(R) = \sum_j \left[
        \underbrace{
            f_{\mu}(|R_{ij}|, z_i, z_j)
        }_{\text{radial}}
        \cdot 
        \underbrace{
            \overbrace{
                R_{ij} \otimes \cdots \otimes R_{ij}
            }^{\nu \text{ times}}
        }_{\text{angular}}
\right]
\end{equation}
where the function $f_{\mu}$ are the radial distributions of the local atomic environment around atom $i$, specified to the neighboring atom $j$. The term $R_{ij} \otimes \cdots \otimes R_{ij}$ are tensors of rank $\nu$, encoding the angular information about the local environment. There are two key parameters that determine the accuracy and computational cost of the trained Machine-Learned Force Field (MLFF): the cutoff radius ($R_{\text{cut}}$) and the maximum level ($lev_\text{{max}}$). In this work, we choose $R_{\text{cut}} = 5\, \text{\AA}$ and $lev_\text{{max}} = 20$. The energy, force and stress data are assigned weights of 1, 0.01, and 0.001.

\subsection*{Feature extraction based on the SOAP descriptor and neural network model}
Smooth Overlap of Atomic Positions (SOAP) is a descriptor that encodes regions of atomic geometries by using a local expansion of a Gaussian smeared atomic density with orthonormal functions based on spherical harmonics and radial basis functions\cite{SOAP-1}. We employ the DScribe\cite{Himanen2020} package to extract all the SOAP vectors at distorted tetrahedral (T) sites and saddle (S) points. The parameters $r_{\text{cut}}$ = 7.0 \AA, $n_{max}$ = 8, and $l_{max}$ = 6 are considered, resulting in the SOAP vector with the dimension of 1 × 1536. The input feature is the SOAP difference between the connected T sites and S points. We use Pytorch\cite{Paszke2019} to train neural network models for predicting diffusion barriers from input features. The network is constructed as a sequence of modules, starting with an input layer that takes data of specified size, followed by the several hidden layers. Each hidden layer is a combination of a fully connected (\textit{nn.Linear}) layer and a non-linear activation function (\textit{ReLU}). The output of the final hidden layer is passed to a linear output layer that matches the desired output size.

\subsection*{Kinetic Monte Carlo simulation}
KMC simulations are conducted to study the long timescale H diffusion in elemental metals and MPEAs. In BCC lattice, H diffusion occurs through a tetrahedral (T) site jumping to its nearest neighboring T sites, each of which with a rate defined as $k_i = k_0 \cdot e^{-\left( \text{DB}_i / (k_B \cdot T) \right)}$, where $k_0$, $k_\text{B}$ and $T$ denote the attempt frequency ($1.5 \times 10^{13} \, \text{s}^{-1}$), Boltzmann constant ($8.61733326 \times 10^{-5}$ eV/K), and simulation temperature, respectively. It should be noted that the attempt frequency here ($1.5 \times 10^{13} \, \text{s}^{-1}$) is different from the default value ($1.0 \times 10^{13} \, \text{s}^{-1}$) in traditional KMC simulations. We choose $1.5 \times 10^{13} \, \text{s}^{-1}$ by minimizing the difference of diffusion coefficients ($D$) between KMC and MD for four pure elements. As a result, our KMC simulations consider the temperature effect to some extent, which is evidenced by the comparable diffusion coefficients between KMC and MD across 969 compositions in Fig. S13. $\text{DB}_i$ represents the local diffusion barrier along the jump path $i$, obtained by the neural network model. The total jump rate is the sum of all individual rates, $R = \sum_{i=1}^{4} k_i$, where 4 denotes the four connected T-T paths. To simulate the atomic hydrogen (H) jump, we generate a uniform random number $u$ within the interval (0,1]. We then select a diffusion path $p$ that meets the following condition: $\sum_{i=1}^{p-1} \frac{k_i}{R} \leq u \leq \sum_{i=1}^{p} \frac{k_i}{R}$. The diffusion of H along path is simulated as the H atom progresses towards the next T site at the end of this path. The time scale for such jump is estimated by $t = -\ln(\rho) / R$ with a random number $0 < \rho < 1$. This trajectory of H is analyzed to determine the mean squared displacement (MSD) and diffusion coefficients for each composition. Since the regular KMC (rKMC) and the constrained KMC (cKMC) have the same diffusion trajectory, the only difference between them is the time scale at each step which is simply a function of temperature. For a comprehensive explanation of determining reliable diffusion coefficients from KMC, please refer to Supplementary Note 3. It is important to note the significant challenges in obtaining reliable diffusion coefficients for $\text{Ta}_{20}\text{W}_{80}$ at low temperatures, such as 300 K and 400 K, using KMC methods. These challenges arise primarily due to the “small-barrier problem”, where H becomes trapped in superbasins\cite{Andersen2019}. On the other hand, nuclear quantum effects play a non-negligible role at these relatively low temperatures\cite{Kwon2023}, rendering regular KMC or MD simulations inadequate. Path integral molecular dynamics, are necessary to accurately capture these effects and will be the focus of our future work.

\subsection*{Machine-learning symbolic regression using SISSO}
We apply the Sure Independence Screening and Sparsifying Operator (SISSO) method \cite{Ouyang2018} in our machine-learning symbolic regression analysis. This method targets the effective activation energy ($Q_{\text{VFT}}$) and the Vogel temperature ($T_0$) within the Vogel-Fulcher-Tammann (VFT) model. Our analysis includes fifteen statistical features that characterize the intrinsic diffusion barrier (DB) and solution energy (SE). These are represented across eight features for DB: average (DB$_{\text{ave}}$), standard deviation (DB$_{\text{sd}}$), skewness (DB$_{\text{sk}}$), kurtosis (DB$_{\text{kt}}$), range (DB$_{\text{range}}$), interquartile range (DB$_{\text{IQR}}$, defined as Q3 - Q1 where Q1 and Q3 correspond to the 25th and 75th percentiles, respectively), 5th percentile (DB$_{0.05}$), and 95th percentile (DB$_{0.95}$). Additionally, seven features are used for SE: standard deviation (SE$_{\text{sd}}$), skewness (SE$_{\text{sk}}$), kurtosis (SE$_{\text{kt}}$), range (SE$_{\text{range}}$), interquartile range (SE$_{\text{IQR}}$), 5th percentile (SE$_{0.05}$), and 95th percentile (SE$_{0.95}$). The feature construction is conducted by applying the operator set [+, -, $\times$, $\div$, $^{-1}$] to the fifteen features. The cross validation of the SISSO approach is shown in Supplementary Note 4 and Fig. S15.

\subsection*{MC/MD simulation}
The MC/MD simulations are conducted to generate chemical short range order (SRO) at 300 K by LAMMPS\cite{Thompson2022}. The existing MLFF is used to describe the atomic interactions in MoNbTaW systems\cite{Yin2021}. The samples are initially relaxed and equilibrated at 300 K and zero pressure under the isothermal-isobaric (NPT) ensemble through MD. After that, MC steps consisting of attempted atom swaps are conducted, hybrid with the MD. In each MC step, a swap of one random atom with another random atom of a different type is conducted based on the Metropolis algorithm in the canonical ensemble. 100 MC swaps are conducted at every 1000 MD steps with a time step of 0.001 ps during the simulation. $3\times10^6$ steps are conducted in MC/MD simulations for all compositions. OVITO is used to visualize atomistic structures\cite{Stukowski2010}.

\section*{Data availability}
The data that support the findings of this study are available upon reasonable request. The parameter files of the potential and training dataset will be published upon the publication of this paper.

\section*{Code availability}
All simulations are executed using open-source software LAMMPS. The machine learning force field was trained and validated by the MLIP package\cite{Novikov2021}. All source codes of the NN-KMC are available at the GitHub repository: \href{https://github.com/ufsf/H-diffusion-in-BCC-MPEAs}{https://github.com/ufsf/H-diffusion-in-BCC-MPEAs}.

\section*{Acknowledgments}
This work was sponsored by Nederlandse Organisatie voor WetenschappelijkOnderzoek (The Netherlands Organization for Scientific Research, NWO) domain Science for the use of supercomputer facilities. The authors also acknowledge the use of DelftBlue supercomputer, provided by Delft High Performance Computing Center (https://www.tudelft.nl/dhpc).

\section*{Author Contributions}
F.S. performed high-throughput DFT calculations, developed the machine learning force field, performed the atomistic simulations, and wrote the first draft of the manuscript, F.S. and Z.W. developed the artificial neural network model, F.S., L.L. and P.D. conceptualized the project and designed the research. F.S., L.L., P.D., W.G., Y.J. and C.D. analyzed the simulation data. All the authors contributed to the interpretation of the data.

\section*{Conflict of Interest}
The authors declare no conflict of interest.

\bibliography{ref}

\newpage
\setcounter{figure}{0}
\renewcommand{\thefigure}{S\arabic{figure}}

\section*{Supplementary Material}
\subsection*{Supplementary Note 1. MoNbTaW-H MLFF validation}
We conduct four validation tasks to confirm the broad applicability of our developed machine-learning force field (MLFF) in accurately modeling MPEA-H interactions across the entire compositional space, including the effects of chemical short-range order (SRO).

In Task 1, we generate 1,000 random quaternary 54-atom configurations with compositions ranging from 1.85\% to 98.15\%, each containing a single randomly inserted H atom. As shown in Fig. \ref{figs2}d, there is remarkable agreement between the MLFF and DFT results, with a mean absolute error (MAE) of just 2.38 meV/atom across various compositions.

For Task 2, we introduce SRO using hybrid Monte Carlo/Molecular Dynamics (MC/MD) simulations at 300 K and 700 K in a 36-atom equimolar $\text{Mo}_{25}\text{Nb}_{25}\text{Ta}_{25}\text{W}_{25}$ system. Subsequently, a single H atom is randomly inserted into tetrahedral (T) sites within the configurations exhibiting different SRO. Fig. \ref{figs2}e demonstrates the MLFF's impressive accuracy in predicting the H solution energy at T sites, with an MAE of 19.24 meV compared to DFT calculations.

Task 3 involves examining the H solution energy at 216 T sites and performing climbing image nudged elastic band (CI-NEB) calculations for all 432 T-T H diffusion paths within a 36-atom $\text{Mo}_{5.56}\text{Nb}_{33.33}\text{Ta}_{16.67}\text{W}_{44.44}$ system, as shown in Fig. \ref{fig2}f. The MAE for H solution energy between the MLFF and DFT is only 13.32 meV, demonstrating the MLFF's exceptional accuracy in both static and transition state calculations for non-equimolar systems.

In Task 4, the minimal disparity in H diffusion barriers (MAE = 10.4 meV) between the VASP and LAMMPS results, as depicted in Fig. \ref{figs2}g, further underscores the MLFF's capability for accurate NEB calculations.

It is important to note that the prediction errors in Fig. \ref{figs2}e-g are not normalized by the number of atoms because determining the exact number influencing H solution energy and diffusion barriers is challenging. However, even if we consider only 20 atoms as the influencing factor, the errors remain below 1 meV/atom, which aligns well with our MLFF training and test results.

\newpage
\subsection*{Supplementary Note 2. Lattice distortion-corrected atomic descriptor for diffusion barrier prediction}
To perform H diffusion simulations in a supercell sized $20 \times 20 \times 20$ using Kinetic Monte Carlo (KMC), Fig. \ref{figs7}f shows that over one million diffusion barriers are required as inputs. The efficiency of our machine-learned force field makes it computationally impractical to calculate all these barriers directly using the CI-NEB method. To address this challenge, previous studies employed various machine learning-driven KMC (ML-KMC) techniques\cite{Xu2023,Xing2024}. The core of ML-KMC involves calculating diffusion barriers cost-effectively, typically via a pre-trained neural network (NN) model that utilizes local atomic environments (LAE) as inputs. However, previous studies only utilized unrelaxed LAE, while the lattice distortion effects are ignored. In this study, we demonstrate that employing relaxed LAE significantly improves the predictive accuracy of NN models compared to using unrelaxed LAE, as evidenced by the results in Figs. \ref{figs4} and \ref{figs5}. We employ the Smooth Overlap of Atomic Positions (SOAP) descriptor\cite{SOAP-1} to quantify the metallic environment around a H atom. The difference between the SOAP vectors of the unrelaxed saddle site ($\text{S}_0$) and the initial tetrahedral site ($\text{T}_0$) is employed to represent the forward diffusion path, as depicted in Eq. \ref{eqs1}:
\begin{equation} \label{eqs1} \Delta \text{SOAP}_0 = \text{SOAP}(\text{S}_0) - \text{SOAP}(\text{T}_0) \tag{S1} \end{equation} 
Here, $\text{S}_0$ represents the midpoint between two connected $\text{T}_0$ sites. However, the real T sites and especially the saddle points are unknown due to significant lattice distortion in MPEAs.

To enhance the accuracy of our NN models without substantial computational overhead, we incorporate atomic lattice distortion into our SOAP descriptor (Fig. \ref{figs6}). This involves approximating the distorted T sites and saddle points as outlined in Eqs \ref{eqs2} and \ref{eqs3}:
\begin{equation} \label{eqs2} \mathbf{X}(\text{T}) = \mathbf{X}(\text{T}_0) + \sum_i d\mathbf{X}(\text{T}_i) \tag{S2} \end{equation}

\begin{equation} \label{eqs3} \mathbf{X}(\text{S}) = \mathbf{X}(\text{S}_0) + \sum_i d\mathbf{X}(\text{S}_i) \tag{S3} \end{equation} 
Here, \textbf{X}(\text{T}) and \textbf{X}(\text{S}) denote the predicted positions of T and S sites, respectively, and $d\mathbf{X}(\text{T}_i)$ and $d\mathbf{X}(\text{S}_i)$ represent the atomic displacement vectors of neighboring metal atoms in the first shell of nearest neighbors. The test performance of various NN models using the modified $\Delta \text{SOAP}$ is presented in Fig. \ref{figs4}. Compared with $\Delta \text{SOAP}_0$, the only additional computation for $\Delta \text{SOAP}$ involves structural relaxation of the MPEA matrix by releasing stress and atomic forces, a difference that is largely negligible.

\newpage
\subsection*{Supplementary Note 3. Determination of diffusion coefficients in KMC simulations}
Various methods have been employed in prior studies to determine the diffusion coefficient ($D$). One common approach involves conducting multiple independent diffusion simulations and calculating the averaged mean squared displacement (MSD)\cite{Kwon2023}. Alternatively, some studies performed extended-duration diffusion simulations to achieve a converged MSD\cite{Zhou2016}. Both techniques necessitate precise MSD calculations. In molecular dynamics (MD) simulations of H diffusion in aluminum\cite{Zhou2016}, it has been recommended to use a fixed time interval of approximately 4.4 ps, akin to the residence time, and a total simulation duration of about 10 ns to reliably extract MSD and $D$. However, these parameters are not universally applicable to H diffusion in other materials due to varying diffusion barriers. This challenge is compounded by the potential presence of strong trapping sites in MPEAs, resulting in highly non-uniform H diffusion where the residence time remains undetermined, as depicted in Figs \ref{figs14}a and \ref{figs14}b.

To reliably determine MSD and $D$, we propose extracting effective trajectories (illustrated as red lines in Fig. \ref{figs14}c) from the complete trajectories (blue lines in Fig. \ref{figs14}c), using a jump distance-based criterion. The critical parameter here is the critical distance ($d_c$), where only trajectory points separated by more than $d_c$ are considered to represent an effective jump. Fig. \ref{figs14}d displays the $d_c$-dependent $D$ derived from the same H diffusion trajectory shown in Fig. \ref{figs14}b. We observe that a small $d_c$ (2 \AA) significantly overestimates $D$. This overestimation occurs because using a small $d_c$ results in the misinterpretation of repetitive jumping within trapping sites as effective diffusion. Conversely, $D$ decreases with increasing $d_c$, eventually stabilizing at a plateau for $d_c$ = 20 \AA, which corresponds to the characterized size of the trapping sites. Consequently, we employ $d_c$ = 20 \AA in our study.

\newpage

\subsection*{Supplementary Note 4. Cross-validation of machine-learning symbolic regression models}
\subsubsection*{Cross-validation for the Vogel temperature $T_0$}
The performance of the SISSO model is rigorously evaluated using both 5-fold and 10-fold cross-validation to assess its predictive capability and check for potential overfitting. In the 5-fold cross-validation (Fig. \ref{figs15}a), the model exhibited high training coefficients of determination ($R^2$), with values ranging from 0.9418 to 0.9482 and a mean $R^2$ of 0.9447. The corresponding test $R^2$ values were also robust, ranging from 0.9211 to 0.9563 with a mean of 0.9358. The minimal difference between the training and test $R^2$ values (mean difference of 0.0089) indicates excellent generalization performance and suggests that the model is not overfitting to the training data. Similarly, in the 10-fold cross-validation (Fig. \ref{figs15}b), the training $R^2$ values remained consistently high, ranging from 0.9398 to 0.9478 with a mean of 0.9438. The test $R^2$ values demonstrated strong predictive accuracy, ranging from 0.8762 to 0.9640 and averaging at 0.9368. Although one fold exhibited a slightly lower test $R^2$ of 0.8762, the overall consistency of the test $R^2$ values and the low standard deviation (approximately 0.0261) reinforce the model's reliability and robustness. The small mean difference between training and test $R^2$ values in the 10-fold cross-validation (0.0070) further confirms the absence of overfitting. These results collectively indicate that the SISSO model maintains high predictive performance across different subsets of the data, with consistent $R^2$ values and minimal variability between folds. The strong agreement between training and test $R^2$ values demonstrates the model's ability to generalize well to unseen data. Therefore, we confidently proceeded to use the entire dataset to derive the final analytical expressions for $T_0$ based on the statistical properties of intrinsic DB and SE spectra.

\subsubsection*{Cross-validation for the effective activation energy $Q_\text{VFT}$}
In the 5-fold cross-validation (Fig. \ref{figs15}c), the training coefficients of determination ($R^2$) ranged from 0.9133 to 0.9272, with a mean $R^2$ of 0.9205 and a standard deviation of approximately 0.0051. The corresponding test $R^2$ values varied from 0.8755 to 0.9451, yielding a mean of 0.9098 and a standard deviation of approximately 0.0246. The small difference between the mean training and test $R^2$ values (0.0107) indicates that the model generalizes well to unseen data, though there is some variability in performance across folds. Similarly, in the 10-fold cross-validation (Fig. \ref{figs15}d), the training $R^2$ values remained consistently high, ranging from 0.9144 to 0.9252 with a mean of 0.9198 and a standard deviation of approximately 0.0035. The test $R^2$ values ranged from 0.8393 to 0.9544, averaging at 0.9121 with a standard deviation of approximately 0.0397. While the mean difference between training and test $R^2$ values was small (0.0077), the higher standard deviation in the test $R^2$ suggests greater variability in the model's predictive performance across different validation sets. These results demonstrate that the SISSO model maintains strong predictive capabilities, with both training and test $R^2$ values generally above 0.91. The minimal differences between training and test $R^2$ values indicate that overfitting is not a significant concern. However, the observed variability in test $R^2$ values, particularly in the 10-fold cross-validation, highlights the importance of comprehensive cross-validation to capture the full range of data variability. Despite these fluctuations, the overall high $R^2$ values affirm the model's robustness and its ability to capture the underlying patterns in the data. Therefore, we proceeded to utilize the entire dataset to derive the final analytical expressions for $Q_\text{VFT}$ based on the statistical properties of the intrinsic DB and SE spectra.

\newpage

\section*{Supplementary Figures}

\begin{figure}[H]
\centering
\includegraphics[width=1\linewidth]{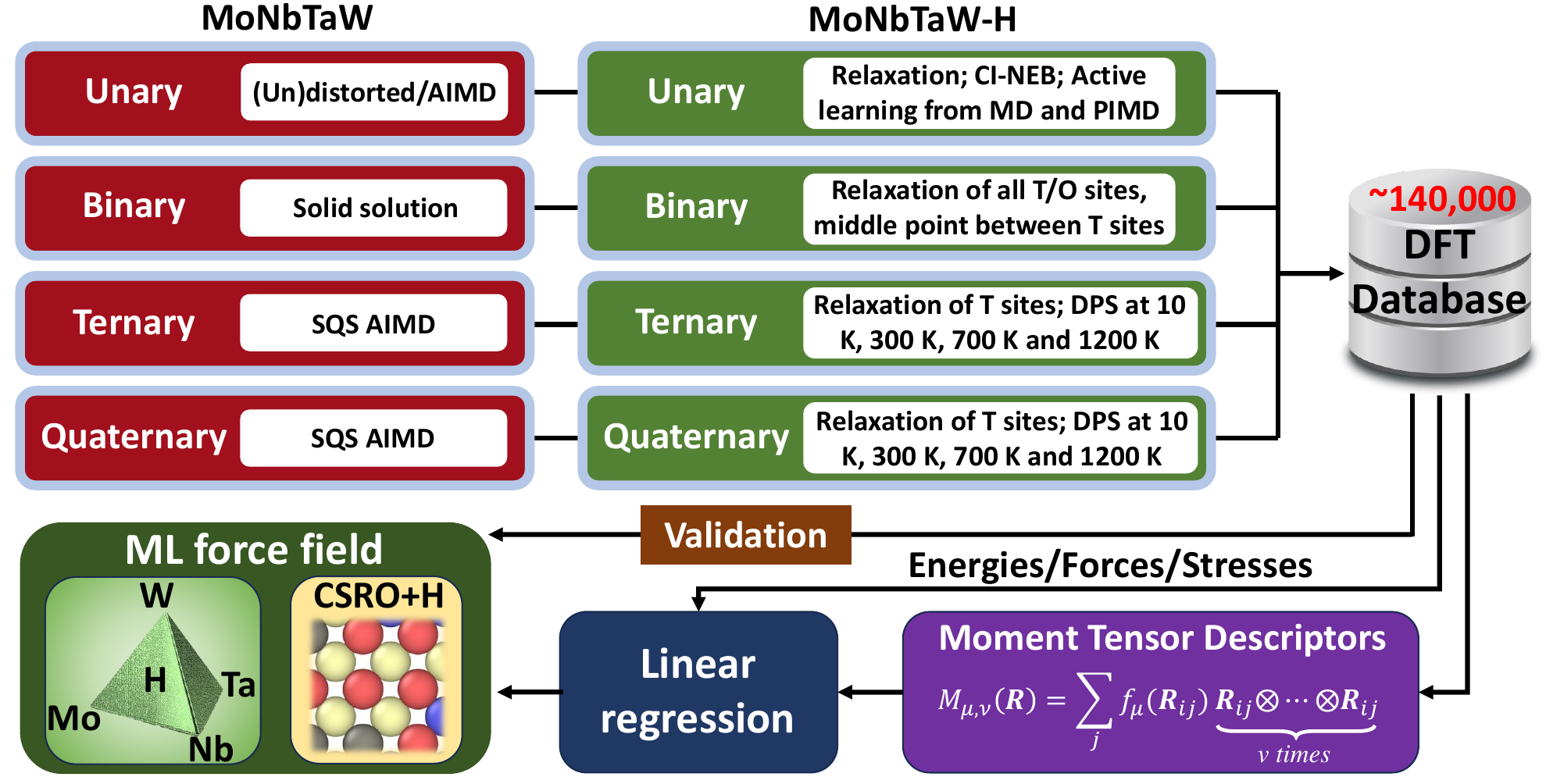} 
\caption{\textbf{Database construction and machine-learning force field development for MoNbTaW-H.} AIMD denotes ab initio molecular dynamics. SQS represents special quasirandom structures. PIMD stands for path integral molecular dynamics. DPS refers to D-optimality-based pre-selection, detailed in the "Methods" section of the main text.}
\label{figs1}
\end{figure}

\begin{figure}
\centering
\includegraphics[width=1\linewidth]{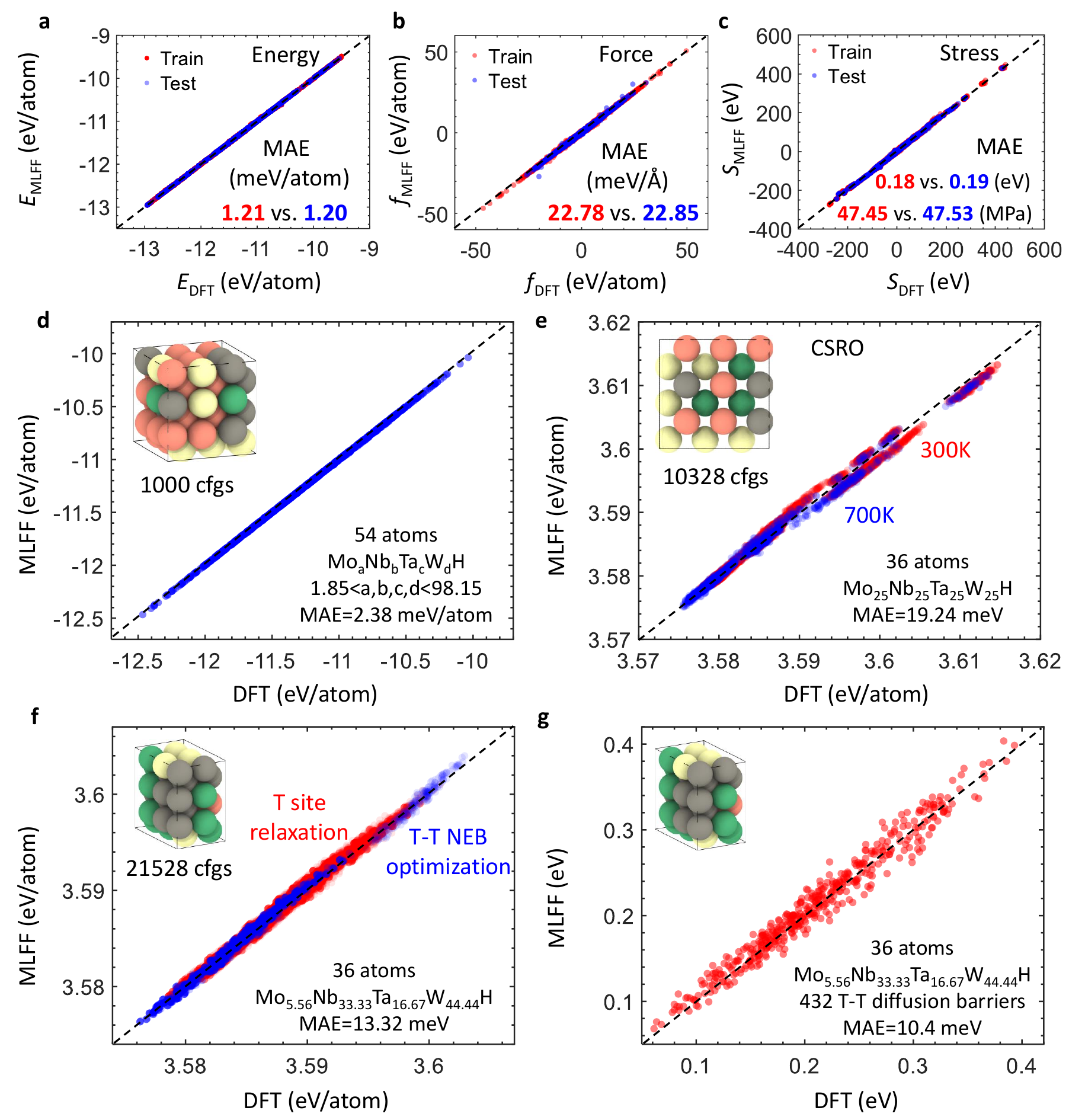} 
\caption{\textbf{Machine-learning force field training, test, and validation.} \textbf{a}-\textbf{c} show training and test errors for potential energies, atomic forces, and stresses. \textbf{d} Validation Task 1: coverage across the entire compositional space illustrated using a random 54-atom configuration with a randomly inserted H atom. \textbf{e} Validation Task 2: H solution energies at tetrahedral (T) sites in 36-atom MoNbTaW configurations with chemical short-range order (SRO) at 300K and 700K. \textbf{f} Validation Task 3: full relaxation with H at all T sites and the climbing image nudged elastic band (CI-NEB) optimization of T-T diffusion paths in non-equimolar $\text{Mo}_{5.56}\text{Nb}_{33.33}\text{Ta}_{16.67}\text{W}_{44.44}$. \textbf{g} Validation Task 4: diffusion barriers of 432 T-T paths in $\text{Mo}_{5.56}\text{Nb}_{33.33}\text{Ta}_{16.67}\text{W}_{44.44}$, obtained using VASP and LAMMPS with the developed MLFF.}
\label{figs2}
\end{figure}

\begin{figure}
\centering
\includegraphics[width=1\linewidth]{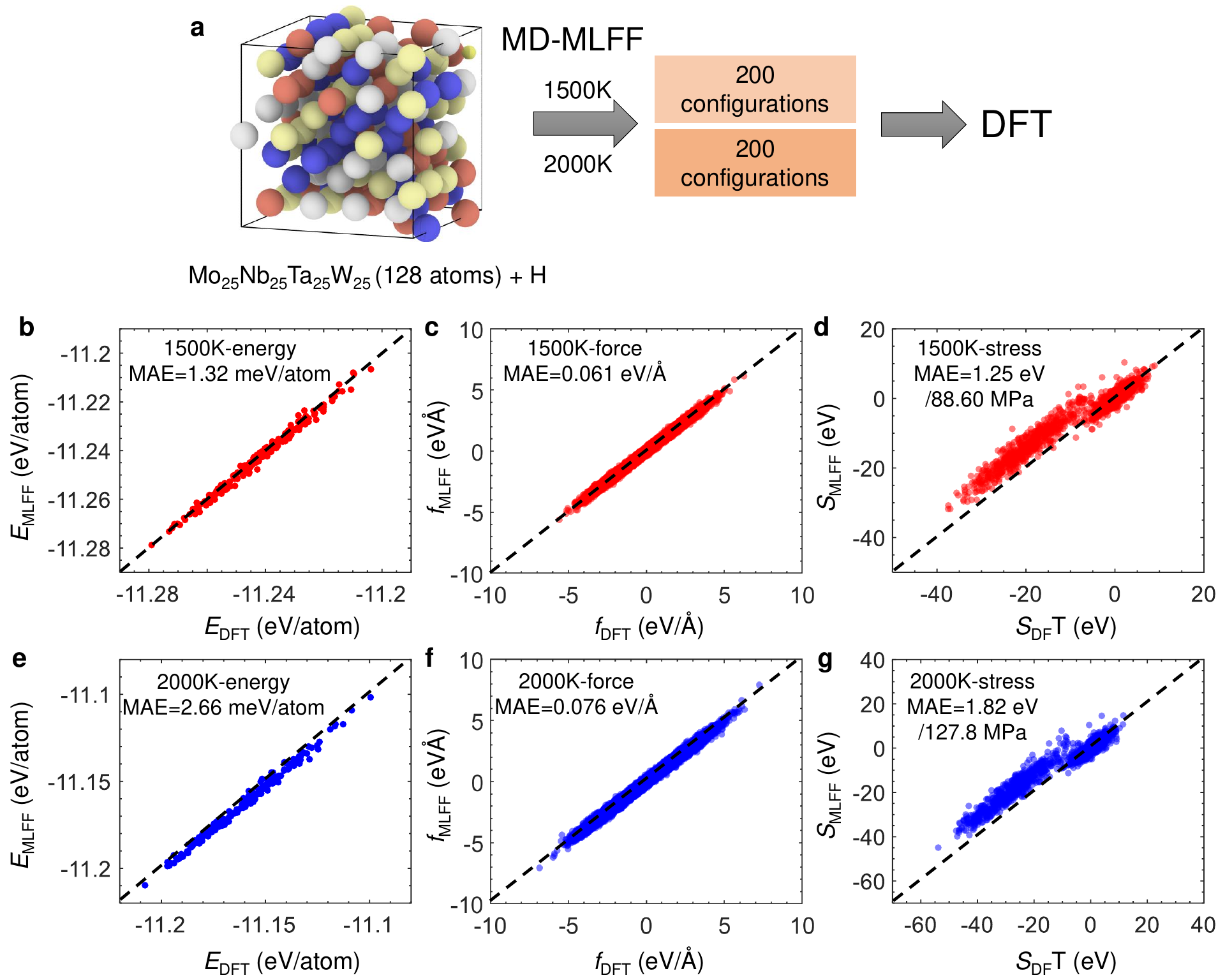} 
\caption{\textbf{Machine-learning force field (MLFF) validation at high temperatures.} \textbf{a} Atomistic configurations and simulation processes used to generate the DFT dataset for high temperature validation. \textbf{b}-\textbf{g} Energy, force, and stress validations using the current MLFF at 1500 K and 2000 K.}
\label{figs3}
\end{figure}

\begin{figure}
\centering
\includegraphics[width=1\linewidth]{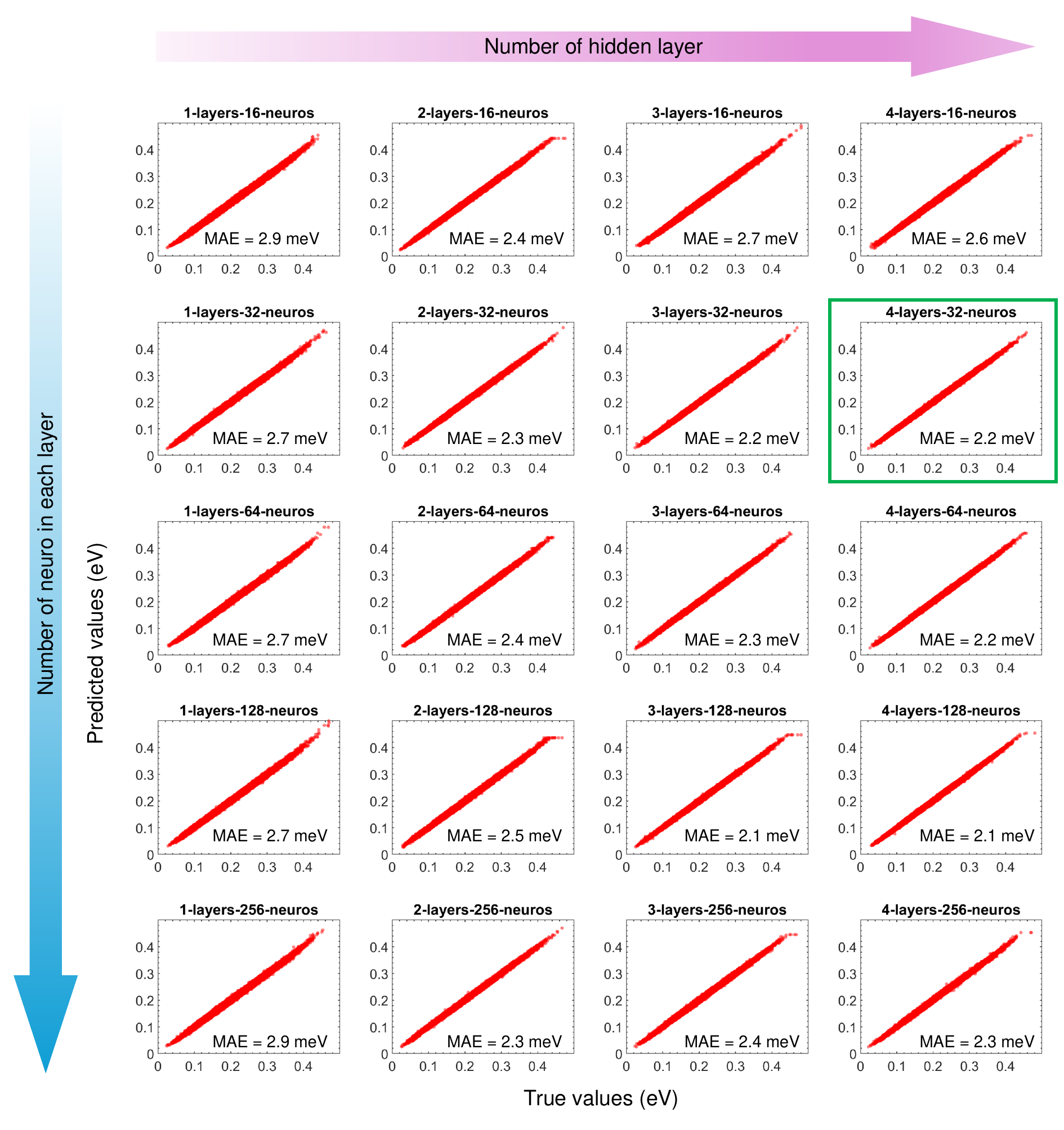} 
\caption{\textbf{Neural network models with different layers using $\Delta \text{SOAP}$ obtained from relaxed structures.}  20 neural network models are trained by varying the number of hidden layer and the number of neuros at each layer. 4-layers-32-neuros indicated by the green rectangular is selected as the final model.}
\label{figs4}
\end{figure}

\begin{figure}
\centering
\includegraphics[width=1\linewidth]{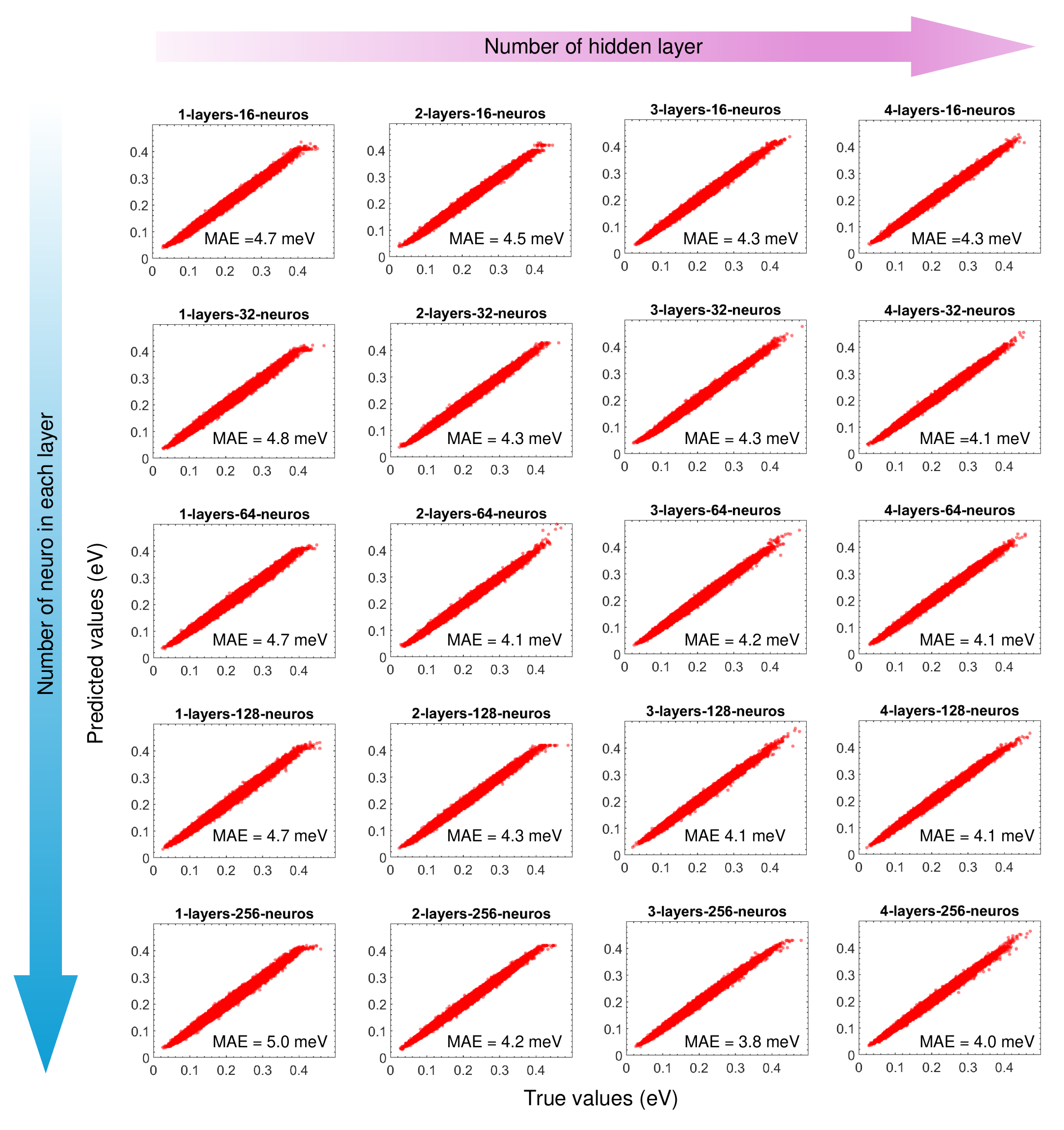} 
\caption{\textbf{Neural network models with different layers using $\Delta \text{SOAP}_0$ obtained from unrelaxed structures.} Twenty neural network models are trained by varying the number of hidden layers and the number of neurons in each layer.}
\label{figs5}
\end{figure}

\begin{figure}
\centering
\includegraphics[width=0.6\linewidth]{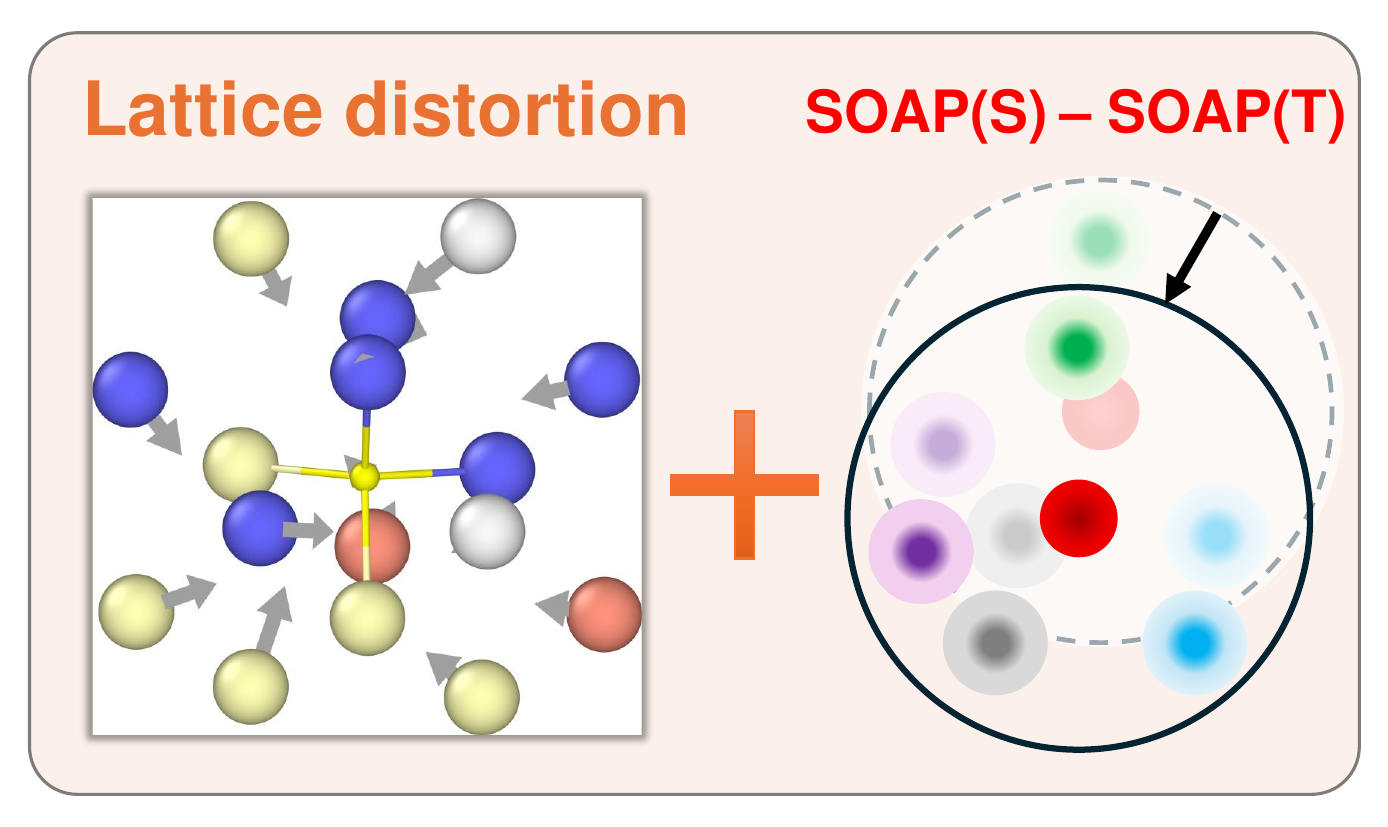} 
\caption{\textbf{Lattice distortion-corrected atomic descriptor used in the neural network model.}}
\label{figs6}
\end{figure}

\begin{figure}
\centering
\includegraphics[width=1\linewidth]{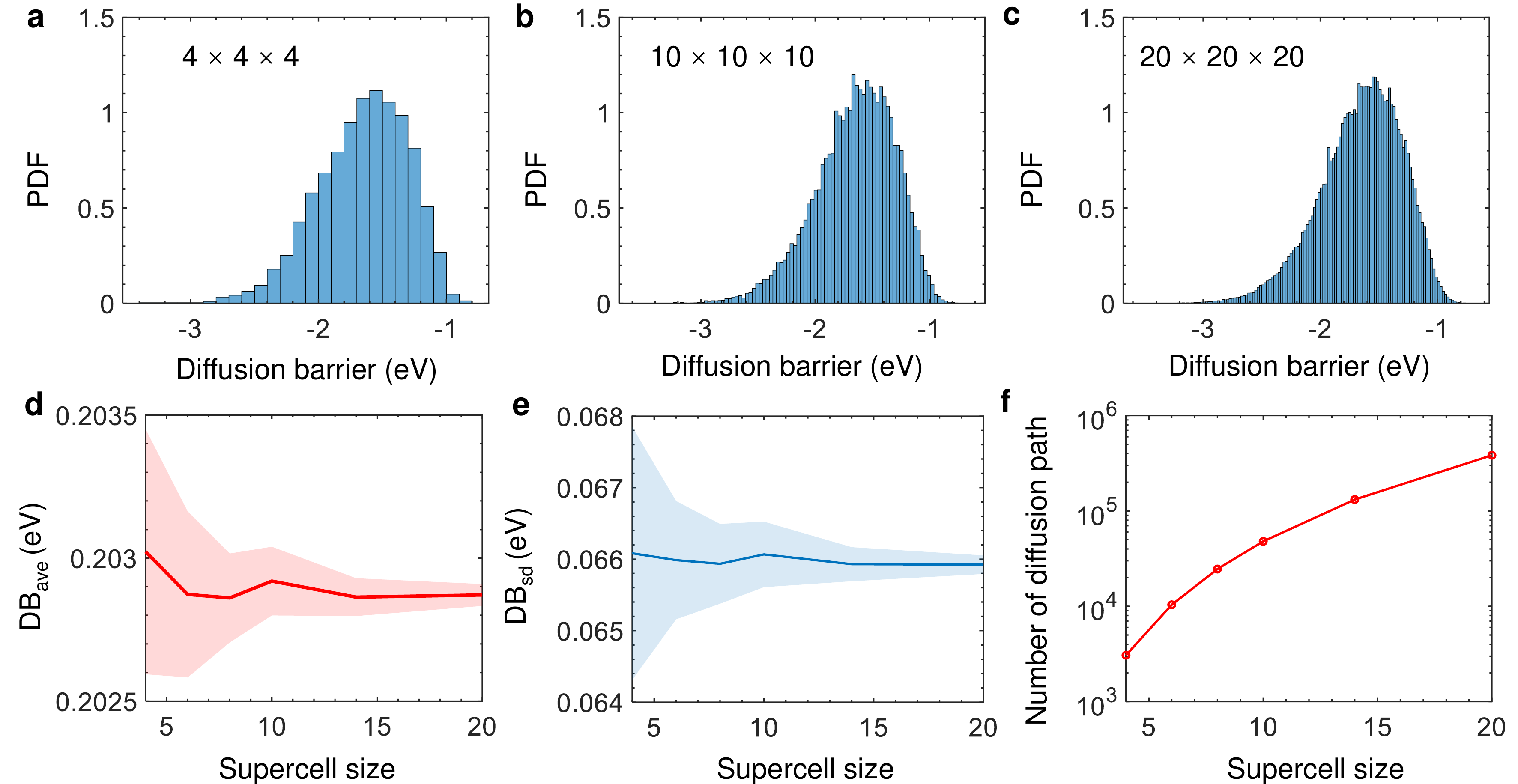} 
\caption{\textbf{Size-dependent diffusion barrier (DB) spectrum in $\text{Mo}_{25}\text{Nb}_{25}\text{Ta}_{25}\text{W}_{25}$.} \textbf{a}-\textbf{c} DB distribution using supercell sizes of $4 \times 4 \times 4$, $10 \times 10 \times 10$, and $20 \times 20 \times 20$. \textbf{d}-\textbf{f} Variations in $\text{DB}_\text{ave}$, $\text{DB}_\text{sd}$, and the number of diffusion paths with respect to supercell size. Figure \textbf{f} reveals that a single supercell of dimensions $20 \times 20 \times 20$ contains approximately half a million diffusion paths.}
\label{figs7}
\end{figure}

\begin{figure}
\centering
\includegraphics[width=1\linewidth]{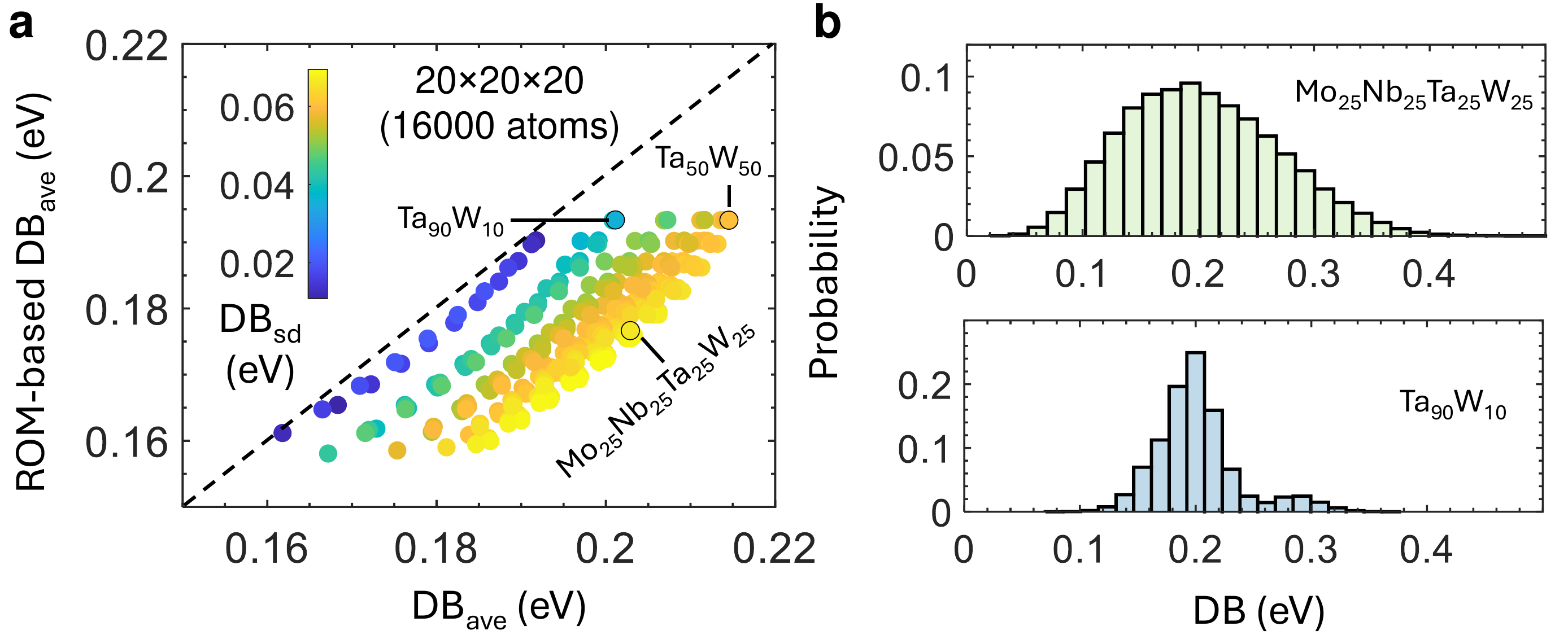} 
\caption{\textbf{Application of the NN model by computing all diffusion barriers for 287 compositions in a $20 \times 20 \times 20$ supercell.} \textbf{a} Mean diffusion barriers ($\text{DB}_\text{ave}$) for each composition compared against rule of mixing (ROM)-based predictions, with standard deviations ($\text{DB}_\text{sd}$) illustrated by different colors for each composition. \textbf{b} Statistical distribution of diffusion barriers for $\text{Mo}_{25}\text{Nb}_{25}\text{Ta}_{25}\text{W}_{25}$ and $\text{Ta}_{90}\text{W}_{10}$.}
\label{figs8}
\end{figure}

\begin{figure}
\centering
\includegraphics[width=1\linewidth]{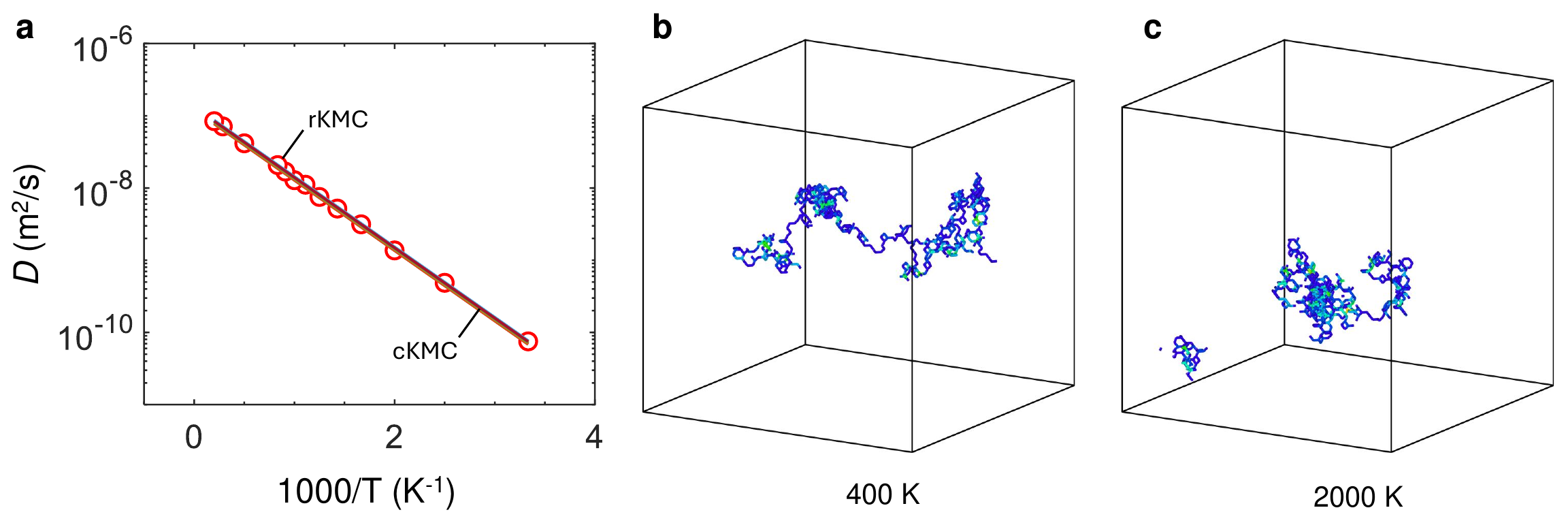} 
\caption{\textbf{Regular KMC (rKMC) and constrained KMC (cKMC) simulation results for pure W.} \textbf{a} Diffusion coefficients at various temperatures, with red circular symbols representing rKMC results and solid lines depicting cKMC results. Both methods predict identical diffusion coefficients at the same temperatures. \textbf{b}, \textbf{c} Hydrogen diffusion trajectories at 400 K and 2000 K, respectively.}

\label{figs9}
\end{figure}

\begin{figure}
\centering
\includegraphics[width=1\linewidth]{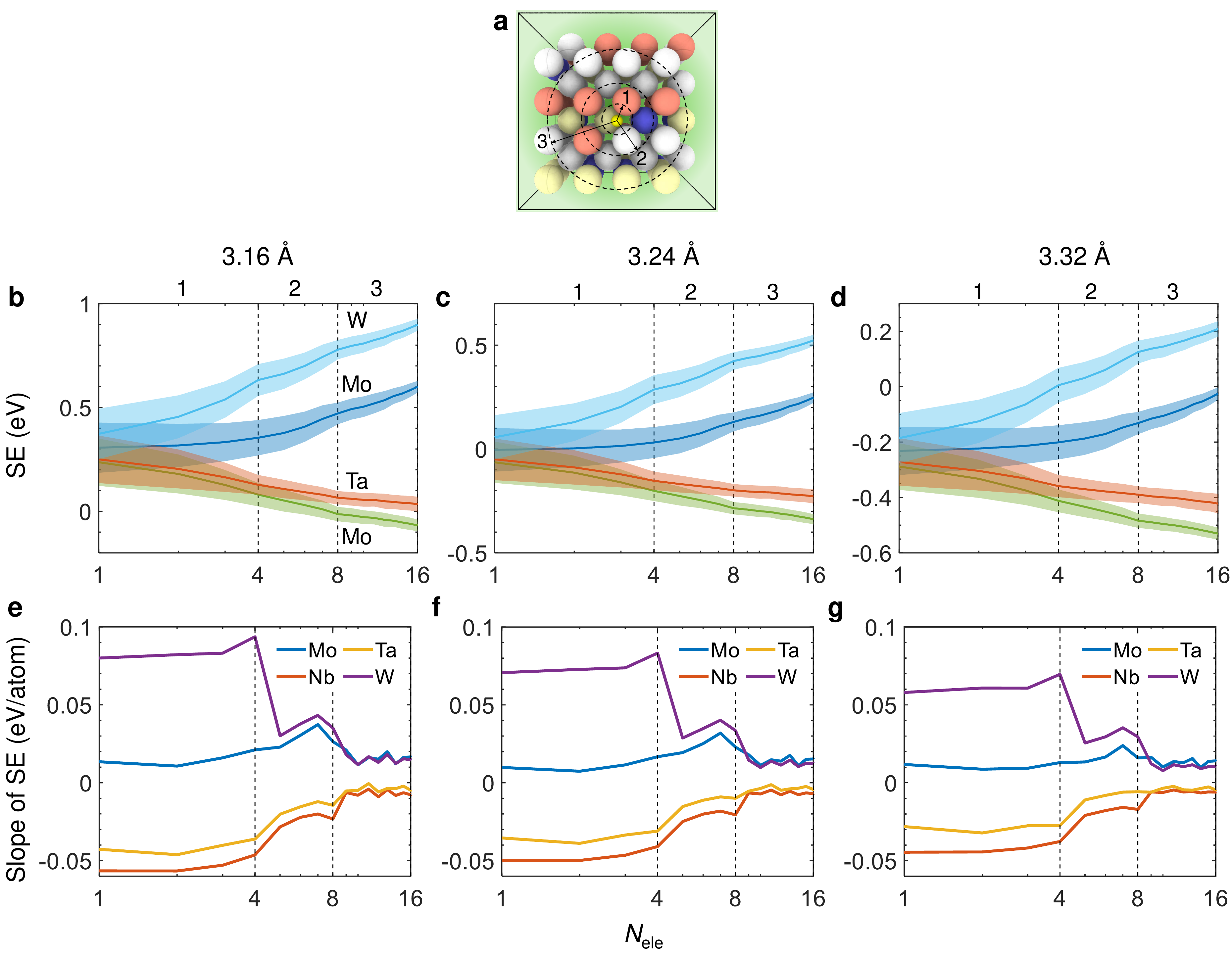} 
\caption{\textbf{The general effects of metallic environments on H solution energy.} \textbf{a} Schematic illustrating the calculation of H solution energy in various chemical environments. \textbf{b}-\textbf{d} Variation of H solution energy with different numbers of a specific element in the nearest neighbors, considering three lattice constants: 3.16 \AA, 3.24 \AA, and 3.32 \AA. \textbf{e}-\textbf{g} The slope of H solution energy as a function of the number of specific elements for different lattice constants.}
\label{figs10}
\end{figure}

\begin{figure}
\centering
\includegraphics[width=1\linewidth]{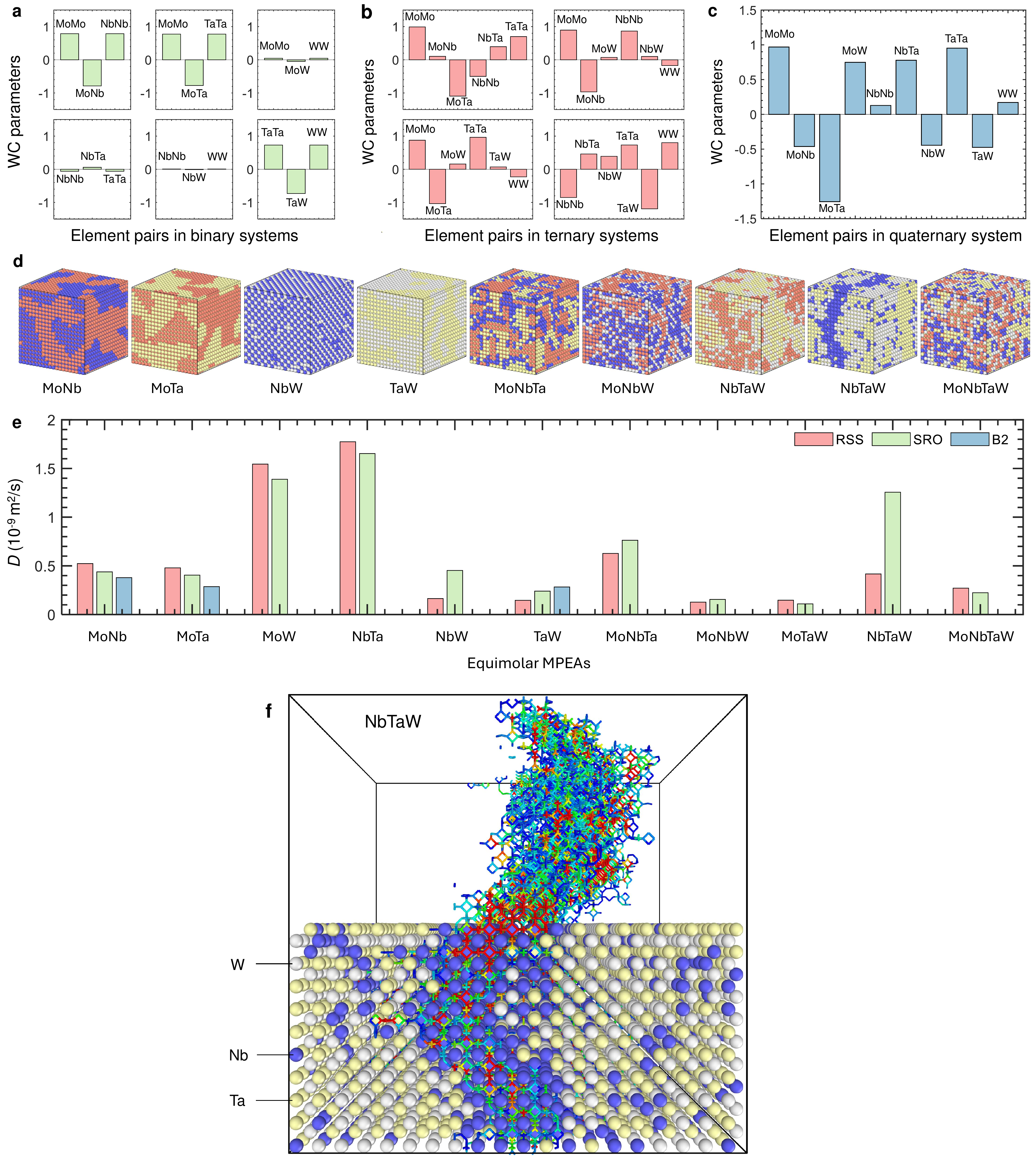} 
\caption{\textbf{The effects of chemical short-range order (SRO) on H diffusion in 11 equimolar multi-principal element alloys (MPEAs).} \textbf{a}-\textbf{c} Warren-Cowley (WC) parameters for all equimolar binary, ternary, and quaternary MPEAs. \textbf{d} Atomic configurations of $20 \times 20 \times 20$ supercells for various compositions, where red, blue, yellow, and white particles represent Mo, Nb, Ta, and W atoms, respectively. For MoNb, MoTa, and TaW, H diffusion is also simulated in configurations with long-range B2 structures. \textbf{e} Diffusion coefficients at 300 K for different compositions, with and without SRO. \textbf{f} H diffusion in the Nb-segregated region of NbTaW with SRO. SRO is observed to have negligible effects on H diffusion in equimolar MPEAs, except in NbTaW, where strong Nb-Nb attractions lead to the formation of extensive Nb-rich regions, facilitating rapid diffusion channels for H.}
\label{figs11}
\end{figure}

\begin{figure}
\centering
\includegraphics[width=1\linewidth]{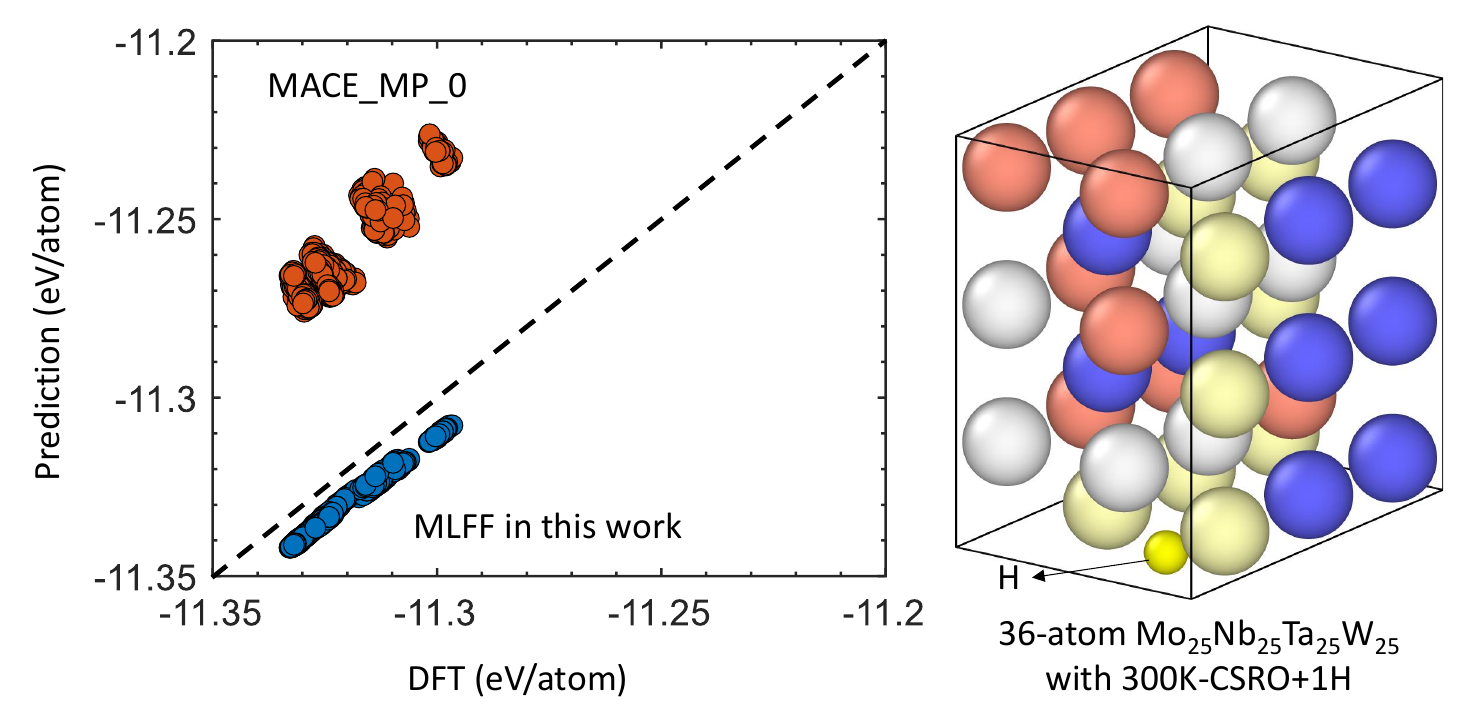} 
\caption{\textbf{Assessment of the latest universal machine learning interatomic potential (uMLIP), MACE-MP-0\cite{Batatia2023}.}}
\label{figs12}
\end{figure}

\begin{figure}
\centering
\includegraphics[width=1\linewidth]{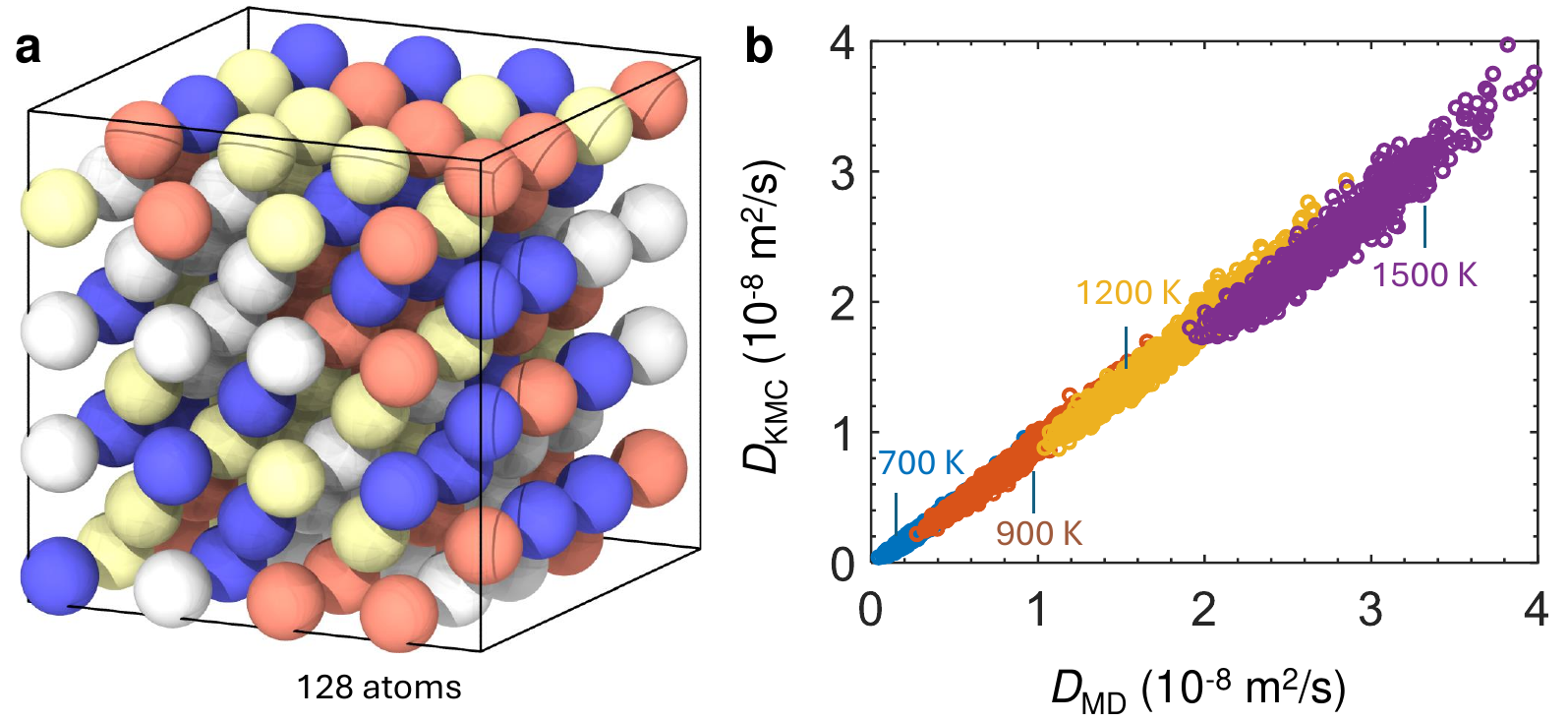} 
\caption{\textbf{Comparison of H diffusion simulations between Kinetic Monte Carlo (KMC) and Molecular Dynamics (MD).} \textbf{a} The $4 \times 4 \times 4$ supercell used for H diffusion simulations in both KMC and MD. \textbf{b} Diffusion coefficients at various temperatures for the two methods. A total of 969 compositions are simulated with a concentration interval of 6.25\% for each element, covering the entire compositional space of MoNbTaW MPEAs. The simulation results from KMC are observed to be in good agreement with those from MD across a wide range of temperatures.}
\label{figs13}
\end{figure}

\begin{figure}
\centering
\includegraphics[width=1\linewidth]{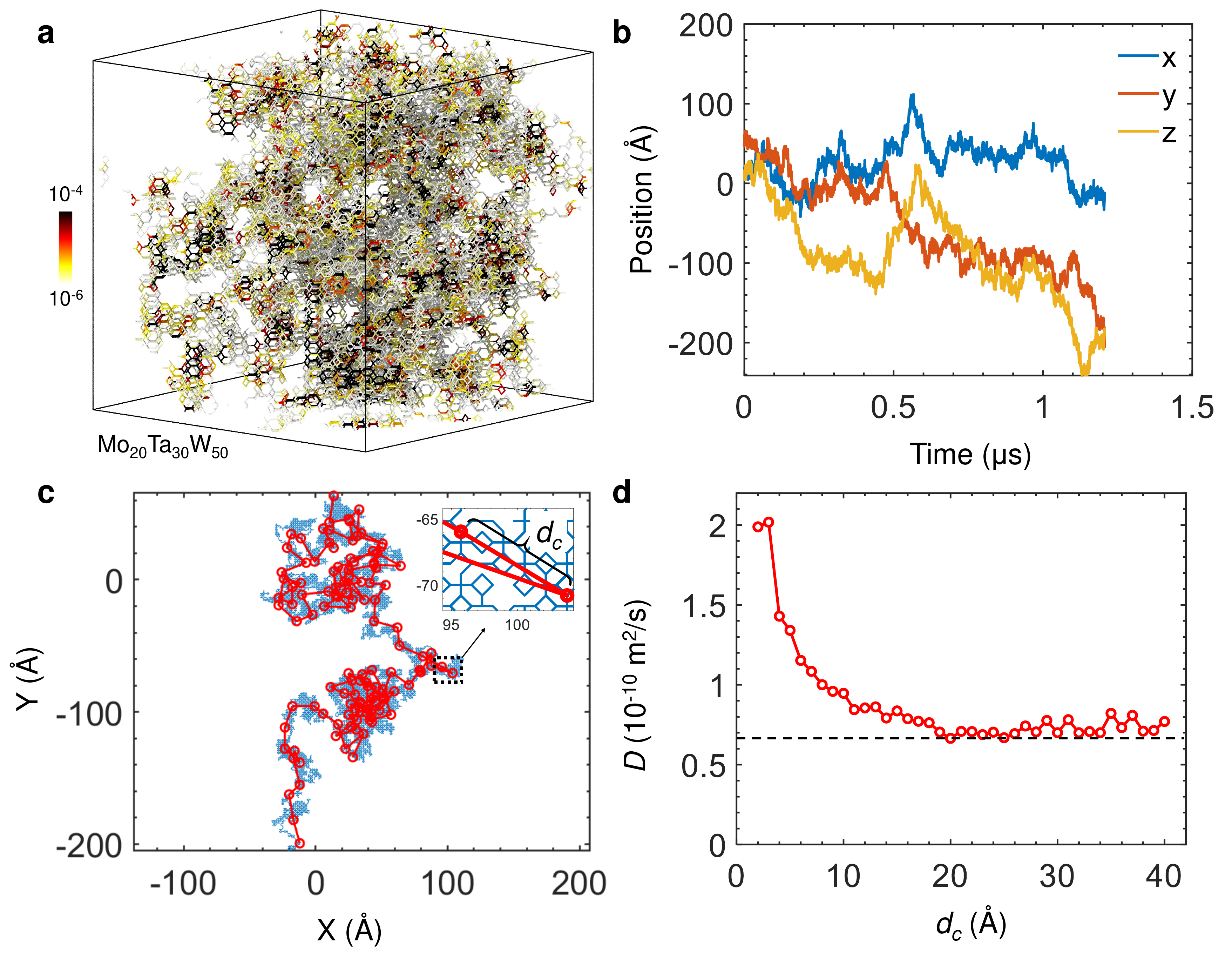} 
\caption{\textbf{Diffusion coefficients determination in KMC simulations.} \textbf{a} Visualization of H visitation frequencies at all accessible tetrahedral (T) interstitial sites at 500 K in $\text{Mo}_{20}\text{Ta}_{30}\text{W}_{50}$ with chemical short-range order (SRO). \textbf{b} Evolution of H atom positions along the $x$, $y$, and $z$ directions. \textbf{c} Full trajectory of H atoms (blue lines) and the extracted trajectory (red lines with markers) in the $x$-$y$ plane, using a critical distance $d_c$ of 20 \AA\ to determine effective jumps. \textbf{d} Diffusion coefficient ($D$) dependency on the critical distance $d_c$.}
\label{figs14}
\end{figure}

\begin{figure}
\centering
\includegraphics[width=1\linewidth]{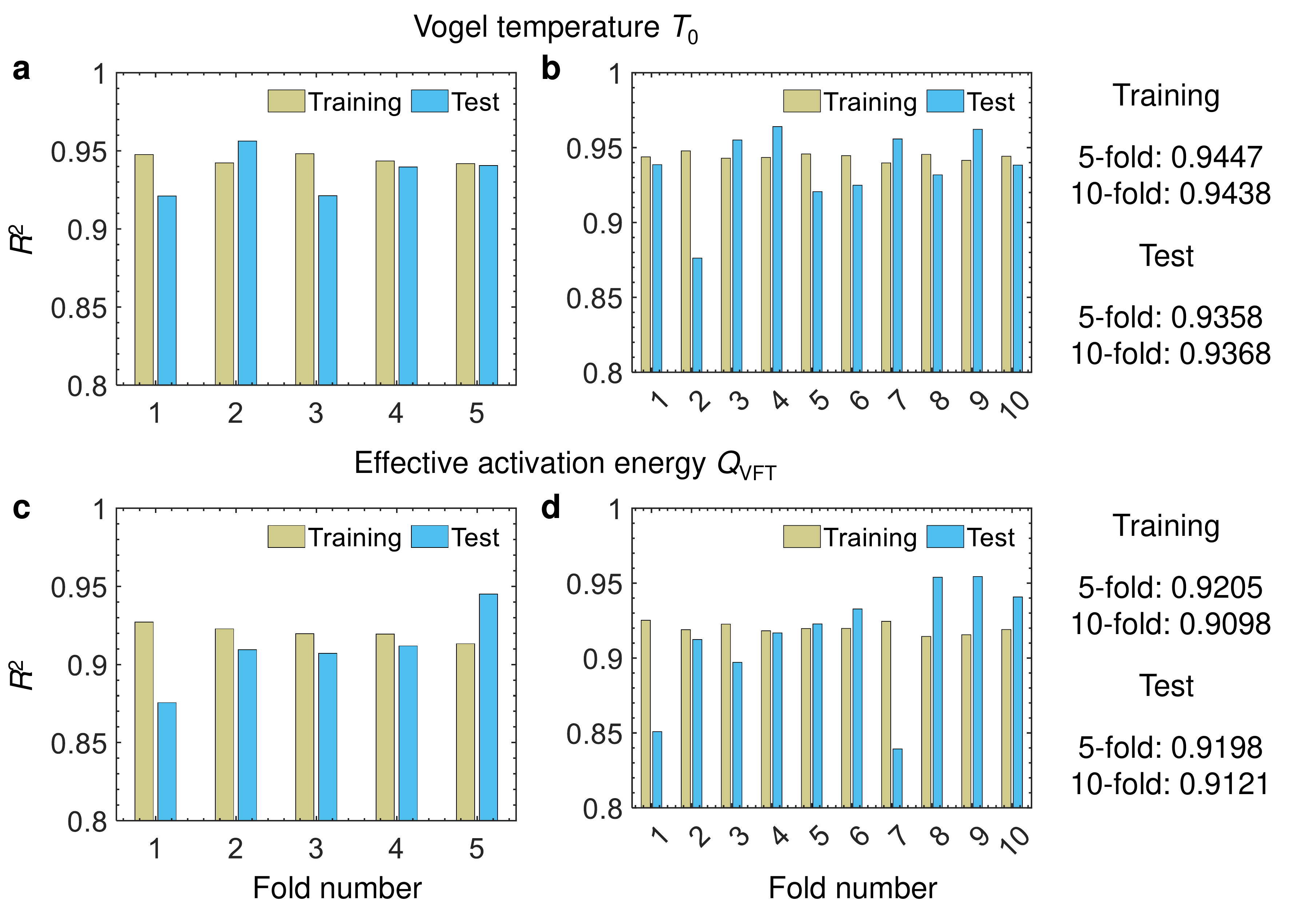} 
\caption{\textbf{Cross-validation for machine-learning symbolic regression models.} \textbf{a}, \textbf{b} 5-fold and 10-fold cross-validation for the Vogel temperature $T_0$. \textbf{c}, \textbf{d} 5-fold and 10-fold cross-validation for the effective activation energy $Q_\text{VFT}$.}
\label{figs15}
\end{figure}

\end{document}